\documentclass[twocolumn,usenatbib]{mn2e}

\usepackage{amsfonts}
\usepackage{amsmath}
\usepackage{amssymb}
\usepackage{aas_macros}
\usepackage{graphicx}
\usepackage{color}
\usepackage{hyperref}
\usepackage{float}
\usepackage{natbib}

\hypersetup{breaklinks,colorlinks,citecolor=blue}

\def\lsim{~\rlap{$<$}{\lower 1.0ex\hbox{$\sim$}}}
\def\bsim{~\rlap{$>$}{\lower 1.0ex\hbox{$\sim$}}}

\def\hmsun{\ {\rm M_\odot/{\it h}}}

\def\dc{\delta_c}

\def\la{\langle}
\def\ra{\rangle}
\def\dd{{\rm d}}
\def\ln{{\rm ln}}

\def\mathbi#1{\textbf{\em #1}}

\def\kvh{\mathrm{\hat{\mathbi{k}}}}
\def\nvh{\mathrm{\hat{\mathbi{n}}}}

\def\nh{\mathit{\hat{n}}}

\def\vk{\mathbi{k}}

\def\vv{\mathbi{v}}
\def\vx{\mathbi{x}}

\def\grad{\mathbi{$\nabla$}}

\newcommand{\fnl}{f_\text{NL}}
\newcommand{\fsky}{f_\text{sky}}
\newcommand{\lmin}{\ell_\text{min}}
\newcommand{\lmax}{\ell_\text{max}}

\def\apjl{Astrophys.\ J.\ Lett.}

\def\aap{Astron.\ Astrophys.}

\def\apjs{Astrophys.\ J. Supp.}

\definecolor{RedWine}{rgb}{0.743,0,0}
\definecolor{RoyalBlue}{rgb}{0.25,.41,.88}
\definecolor{ForestGreen}{rgb}{.13,.54,.13}
\definecolor{DeepPurple}{rgb}{.72,.18,1}

\newcommand{\CIB}{{\mathtt{CIB}}}
\newcommand{\CMB}{{\mathtt{CMB}}}
\newcommand{\FWHM}{{\mathtt{FWHM}}}

\newcommand{\fNL}{{f_{\rm NL}}}
\newcommand{\planck}{{\it Planck}}

\begin{document}


\title[CIB as a window into primordial non-Gaussianities]
{Cosmic Infrared Background anisotropies as a window into primordial non-Gaussianity}

\author[Tucci et al.]
{\parbox[t]{\textwidth}{Marco Tucci\thanks{Email: Marco.Tucci@unige.ch}, 
Vincent Desjacques\thanks{Email: Vincent.Desjacques@unige.ch} and Martin Kunz
\thanks{Email: Martin.Kunz@unige.ch}} \\\,\\
D\'epartement de Physique Th\'eorique and 
Center for Astroparticle Physics (CAP), University of Geneva,  \\
~~~24 quai Ernest Ansermet, CH-1211 Geneva, Switzerland 
}

\date{}
\label{firstpage}
\pagerange{\pageref{firstpage}--\pageref{lastpage}}

\maketitle

\begin{abstract}
The angular power spectrum of the cosmic infrared background (CIB) is a 
sensitive probe of the local primordial bispectrum. 
CIB measurements are integrated over a large volume so that the scale 
dependent bias from the primordial non-Gaussianity leaves a strong 
signal in the CIB power spectrum. Although galactic dust dominates over 
the non-Gaussian CIB signal, it is possible to mitigate the dust 
contamination with enough frequency channels, especially if high 
frequencies such as the \planck\ 857 GHz channel are available. 
We show that, in this case, measurements of the cosmic microwave background 
from future space missions should be able to probe the local bispectrum 
shape down to an amplitude $|\fnl| < 1$.
\end{abstract}

\begin{keywords}
cosmology: theory, large-scale structure of Universe, inflation
\end{keywords}

\section{Introduction}

Cosmology offers powerful probes of the high-energy physics in the early 
Universe.
One of the most important discriminants is primordial non-Gaussianity (PNG), 
which can in principle be tested across our entire past light-cone. 
While standard single-field slow-roll inflationary models generically 
predict only very low levels of PNG, other scenarios can lead to a much 
larger non-Gaussianity in the primordial fluctuations
\citep[for more details see e.g.][]{Bartolo:2004if,Komatsu:2010hc,Chen:2010xka,
Liguori:2010hx,Yadav:2010fz}. Constraints or detections of PNG can thus 
give us insight into the physics at energy scales that are otherwise very 
difficult to access.

The \planck\ satellite \citep{planck_xvii_2015} has already put stringent 
constraints on primordial non-Gaussianity, but there is still much space 
for interesting phenomenology, especially if we can access the regime 
where the amplitude of the PNG is an order of magnitude smaller than the 
current limits.

While the cosmic microwave background (CMB) observations are nearly 
cosmic-variance limited and, therefore, will not improve much over the 
\planck\ constraint, future surveys of the large scale structure (LSS) 
hold the promise of achieving much tighter limits.  
Much effort has already been devoted to constrain PNG from a scale dependence 
in the galaxy bias \citep{Dalal:2007cu, Matarrese:2008nc, Slosar:2008hx}.  
Current LSS limits are at the level of the CMB pre--\planck\ constraints
\citep{Giannantonio:2013uqa, Leistedt:2014zqa}, and they could improve by 
up to $\sim 1$ order of magnitude in the not so distant future
\citep{Agarwal:2013qta,dePutter:2014lna,Raccanelli:2014kga,
Camera:2014bwa,Alonso:2015sfa,Raccanelli:2015oma}. 
Further improvements may come from higher-order statistics such as the galaxy 
bispectrum \citep[see e.g.][]{Scoccimarro:2003wn, Sefusatti:2007ih, Jeong:2009vd},
which encodes much more information on the PNG shape than the non-Gaussian bias.
The latter is predominantly sensitive to the `local' bispectrum shape, which 
peaks on squeezed triangles.

Constraints on $\fnl$ from future measurements of the non-Gaussian bias will 
strongly depend on the survey depth and volume 
\citep{Carbone:2010sb,Hamaus:2011dq,dePutter:2014lna}. 
Unfortunately, even the forthcoming large scale surveys such as Euclid 
\citep{Laureijs:2011gra} or LSST \citep{Abell:2009aa}
will only cover a small fraction of the total comoving volume accessible to 
us. The cosmic infrared background (CIB) traces the LSS over 
a much larger comoving volume and, thus, could potentially outperform future 
galaxy redshift surveys. 
In this paper we show that the CIB indeed is an excellent probe of primordial
non-Gaussianity that allows us, in principle, to reach $|\fnl |<1$ in the 
squeezed bispectrum limit. 
We demonstrate that such an error can be achieved at Fisher matrix level even 
upon taking into account the signal produced by galactic dust, which is much
stronger than the CIB at low multipoles. However, it is essential to have CIB 
maps available at enough frequencies.
 
The paper is organized as follows. We introduce our halo model of the
CIB in \S\ref{sec:model}, and present our forecast for a detection of
the non-Gaussian bias in the CIB angular power spectrum in
\S\ref{sec:fish}. We discuss our findings and, in particular, the
contamination by galactic dust in \S\ref{sec:dust} and
\S\ref{sec:discussion} before concluding in \S\ref{sec:conclusion}.
Throughout the paper, we adopt the standard $\Lambda$CDM cosmological model as 
measured by \planck\ \citep{planck_xvi_2013}:
$\{\Omega_m,\Omega_\Lambda,\Omega_bh^2,\sigma_8,h,n_s\}=
\{0.3183,0.6817,0.02205,0.8347,0.6704,0.9619\}$.

\section{Model of the Cosmic Infrared Background}
\label{sec:model}

\subsection{Halo model of the CIB intensity}

In the usual Newtonian treatment, the observed CIB brightness is given by
\begin{equation}
\label{eq:CIBNewt}
I(\nu,\nvh) =
\int_0^{\chi_*}\!\!\dd z\left(\frac{\dd\chi}{\dd z}\right)\, W_{\nu}^{(\CIB)}(z) 
\,\Big(1 + \delta_\jmath\Big)
\end{equation}
where the integral runs over the (unperturbed) comoving distance $\chi$ along the 
line of sight. It is typically computed up to redshift $z_*\sim 10$, corresponding 
to a comoving distance $\chi_*=\chi(z_*)$, because of the negligible contribution 
of CIB fluctuations at higher redshifts. The redshift weight for CIB fluctuations is
\begin{equation}
\label{eq:WCIB}
W_{\nu}^{(\CIB)}(z)=a(z)\,\bar\jmath_{\nu}(z)\;, 
\end{equation}
where $a(z)$ is the scale factor and $\bar\jmath_{\nu} (z)$ is the mean CIB emissivity 
per comoving volume at frequency $\nu$:
\begin{equation}
\bar\jmath_{\nu}(z)=\int_0^{\bar L_{_{(1+z)\nu}}^\text{\tiny cut}}\!\!\dd L\,
\bar n_g(L,z)\,\frac{L_{(1+z)\nu}}{4\pi}\;.
\label{eq:emod2}
\end{equation}
Here, $\bar n_g(L,z)$ denotes the infrared galaxy luminosity function 
and $L_{(1+z)\nu}$ is the infrared luminosity (in W\,Hz$^{-1}$) at the
rest--frame frequency $(1+z)\nu$. In practice, only faint sources below
the flux detection limit $S_\nu^\text{\tiny cut}$ are included in the CIB, 
such that the integral in Eq.\ (\ref{eq:emod2}) is cut off 
at a (redshift and frequency dependent) luminosity 
$\bar L_{_{(1+z)\nu}}^\text{\tiny cut}(z)$.
The brighter sources are detected and removed from the sky maps.
Finally, $\delta_\jmath\equiv\delta_\jmath(\nu,z,\vx=\chi(z)\nvh)$ is the 
perturbation to the CIB emissivity at frequency $\nu$, redshift $z$ and 
comoving position $\vx=\chi(z)\nvh$ along the line of sight.

Let us focus first on the contribution from the average emissivity.
Following \citet{sha12}, we split the mean emissivity into a sum of two 
contributions,
\begin{equation}
\label{eq:meanj}
\bar\jmath_{\nu}(z)=\int\!\!\dd M\,\bar n_h(M,z)
\Bigl[f_\nu^\text{c}(M,z)+f_\nu^\text{s}(M,z)\Bigr] \;,
\end{equation}
where the average emissivity produced by the central and satellite galaxies
of a given halo at redshift $z$ are 
\begin{align}
f_\nu^\text{c}(M,z) &= \frac{1}{4\pi}N_\text{c}L_{\text{c},(1+z)\nu}(M,z) \\
f_\nu^\text{s}(M,z) &= \frac{1}{4\pi} \int\!\!\dd m\,
\bar n_s(m,z|M)L_{\text{s},(1+z)\nu}(m,z) \;.
\end{align}
In the above expressions, $\bar n_h(M,z)$ and $\bar n_s(m,z|M)$ are the 
halo and sub-halo mass functions, and $M$ and $m$ are the parent halo and 
sub-halo masses. The numbers $N_{\rm c}$ of central
galaxies is specified by the halo occupation distribution
\citep[HOD;][]{ber02,zhe05}. Numerical simulations indicate that
$N_{\rm c}$ typically follows a step-like function \citep{kra04}. We
adopt a characteristic mass $M_\text{cen}$ of the step function of
$3\times10^{11}\hmsun$ \citep{zeh11,des15}, ignoring
any luminosity dependence. In all subsequent calculations, we use the SO 
(Spherical Overdensity) halo and sub-halo mass functions provided by \cite{tin08} 
and \cite{tin10b}, respectively, and integrate the sub-halo mass function
from a minimum halo mass $M_\text{min}=10^{10}\hmsun$ to the parent
halo mass $M$. To identify halos, we adopt a density threshold of 
$\Delta_c=200$ (in unit of the background density $\bar\rho_m(z)$) at all 
redshifts.

In the model used here, the luminosity and clustering of infrared galaxies 
are linked to the host halo mass. The strongly clustered galaxies are 
situated in more massive halos. They typically have more stellar mass and,
therefore, are more luminous. 
Assuming the same luminosity-mass relation for both central and satellite
galaxies, we relate the galaxy infrared luminosity to the host halo
mass through the parametric relation \citep{sha12}
\begin{equation}
L_{(1+z)\nu}(M,z)=L_0\Phi(z)\Sigma(M)\Theta_\CIB[(1+z)\nu]\,.
\end{equation}  
As regards to the luminosity--mass relation, we shall follow the
assumptions used in \citet{planck_xxx_2013}:
\begin{itemize}

\item $L_0$ is an overall normalisation that is constrained from
  measurements of CIB power spectra. In principle, it should be a
  constant but, in order to have a good fit to
  \citet{planck_xxx_2013}, we allow small variations of $L_0$ with the
  frequency, which are at most a 10\,per cent between 217 and
  857\,GHz.

\item The term $\Phi(z)$ describes the redshift-dependence of the
  normalisation. We adopt a power-law scaling
$$
\Phi(z)=(1+z)^{\delta}.
$$

\item We assume a log-normal distribution for the dependence
  $\Sigma(M)$ of the galaxy luminosity on halo mass:
\begin{equation}
\Sigma(M)=\frac{M}{\sqrt{2\pi\sigma^2_\text{L/M}}}
e^{-[\log(M)-\log(M_\text{eff})]^2/2\sigma^2_\text{L/M}}\,,
\end{equation}
where $M_\text{eff}$ characterises the peak of the specific IR
emissivity and $\sigma_\text{M/L}$ describes the range of halo masses
that produces a given luminosity $L$, which is fixed to
$\sigma^2_\text{M/L}=0.5$.

\item For the galaxy spectral energy distribution (SED), we assume a
  modified blackbody shape with a power-law emissivity as in
  \cite{hal10},
\begin{equation}
\Theta_\CIB(\nu)= \left\lbrace\begin{array}{cc}
(\nu/\nu_0)^\beta B_\nu(T_d)/B_{\nu_0}(T_d) & \nu\leq \nu_b \\
(\nu/\nu_0)^{-\gamma} & \nu>\nu_b \end{array}\right. \;,
\label{eq:cibsed}
\end{equation}
where $B_\nu(T)$ is the brightness of a blackbody with temperature $T$
at the frequency $\nu$, and $\nu_0$ is a reference frequency. The dust
temperature $T_d$ is assumed to be a function of redshift according to
$T_d=T_0(1+z)^\alpha$. 
The grey-body and power-law connect smoothly at the frequency $\nu_b$, 
at which the condition $\dd\ln\Theta(\nu,z)/\dd\ln\nu=-\gamma$ is satisfied. 
For our fiducial parameters, $\nu_b$ is always larger than 3000\,GHz, 
and hence the parameter $\gamma$ has small relevance in the frequency 
range we are interested in.

\end{itemize}

For our fiducial halo model parameters, we adopt the values found in
\citet{planck_xxx_2013} by fitting (auto-- and cross--) power
spectra of the CIB anisotropies measured by \planck\ and {\it
  IRAS} at 217, 353, 545, 857 and 3000 GHz (see Table\,\ref{tab:0}).
\begin{table}
\begin{center}
  \caption{Fiducial values of the parameters foud by \citet{planck_xxx_2013} 
    for the luminosity-mass relation of the halo model. \label{tab:0}}
\begin{tabular}{ccccccc}
\hline
$\alpha$ & T$_0$ & $\beta$ & $\gamma$ & $\delta$ & 
$\log(M_\text{eff})$ & L$_0$ \\
& [K] & & & & [M$_{\odot}$] & [W\,Hz$^{-1}$] \\
\hline 
0.36 & 24.4 & 1.75 & 1.7 & 3.6 & 12.6 & 0.95--1.05$\times10^{-3}$ \\
\hline
\end{tabular}
\end{center}
\end{table}

\vspace{3mm}

We now turn to the fluctuations of the CIB brightness across the sky.
In the Limber approximation, the angular power spectrum of CIB fluctuations can be 
written as \citep[e.g.,][]{knox/cooray/etal:2001,son03}:
\begin{align}
C_{\ell}^{(\CIB)}(\nu,\nu') &=\int_0^{z_*}\!\!\dd z\left(\frac{\dd\chi}{\dd z}\right)
\frac{1}{\chi^2}\,W_{\nu}^{(\CIB)}(z)W_{\nu'}^{(\CIB)}(z)
\nonumber \\
&\qquad \times P^{\nu\times\nu'}_{jj}\!\!\left(k=\frac{\ell}{\chi},z\right)\,.
\label{eq:emod1}
\end{align}
Here, $P_{jj}^{\nu\times\nu'}$ is the cross-power spectrum 
$\left\langle\delta_\jmath\delta_\jmath'\right\rangle$ of fluctuations in emissivity 
at frequency $\nu$ and $\nu'$.
$P_{jj}(k,z)$ can be set equal to the 3D power spectrum of galaxies, $P_{\rm gg}(k,z)$, 
under the assumption that fluctuations in the emissivity trace those of the sources.  
In the context of the halo model \citep{sch91,sel00,sco01,coo02}, galaxy power spectra 
are the sum of the contribution of the clustering in one single halo (1--halo term) 
and in two different halos (2--halo term):
\begin{equation}
P^{\nu\times\nu'}_{\rm gg}(k,z)=\bigg. 
P_{\rm gg}^{1h}(k,z,\nu,\nu')+P_{\rm gg}^{2h}(k,z,\nu,\nu')\,,
\end{equation}
where
\begin{align}
P_{\rm gg}^{1h}(k,z,\nu,\nu') & =  \bigg. \frac{1}{\bar{j}_\nu\bar{j}_{\nu'}}
\int\!\!\dd M\,\frac{\dd N}{\dd M}\Bigl[f_\nu^\text{c}(M,z)f_{\nu'}^\text{s}(M,z)
u(k,z|M) \nonumber \\ & \qquad 
+ f_{\nu'}^\text{c}(M,z)f_\nu^\text{s}(M,z) u(k,z|M) \nonumber \\ & \qquad
+ f_\nu^\text{s}(M,z)f_{\nu'}^\text{s}(M,z) u^2(k,z|M) \Bigr] \,,
\nonumber \\
 P_{\rm gg}^{2h}(k,z,\nu,\nu') & = \bigg. \frac{1}{\bar{j}_\nu\bar{j}_{\nu'}}
D_\nu(k,z) D_{\nu'}(k,z) P_\text{lin}(k,z)\,,
\label{eq:emod3}
\end{align}
and
\begin{align}
\label{eq:Dnu}
D_\nu(k,z) &= \int\!\!\dd M\,\frac{\dd N}{\dd M}\,b_1(M,z) u(k,z|M) \\
& \qquad \times \Bigl[f_\nu^\text{c}(M,z)+f_\nu^\text{s}(M,z)\Bigr] 
\nonumber \,.
\end{align}
Here, $u(k,z|M)$ is the normalised Fourier transform of the NFW
density profile \citep{nav97}, $P_\text{lin}(k,z)$ is the linear mass power
spectrum extrapolated to redshift $z$, and $b_1$ is the (Eulerian)
linear halo bias \citep[we use the fitting formula given in][]{tin10a}.
In principle, $P_{jj}(k,z)$ receives also a shot noise contribution, but we 
will ignore it as it is typically of the same order as the 1-halo term.

At very large scales, i.e. at multipole $\ell\lsim10$, the Limber
approximation used in Eq.\ (\ref{eq:emod1}) is not valid. Assuming a
negligible contribution from the 1--halo term, for angular scales
ranging from $\ell=2$ to 40 CIB power spectra can be computed by
\citep{cur15}
\begin{align}
\label{eq:emod6}
C^{(\CIB)}_{\ell}(&\nu,\nu') = \frac{2}{\pi}\int\!\! \dd k\,k^2 \\
& \times \int_0^{z_*}\!\!\dd z\left(\frac{\dd\chi}{\dd z}\right)\,
W_{\nu}^{(\CIB)}(z)j_{\ell}(k\chi(z)) P^{1/2}_{\rm gg}(k,z,\nu) 
\nonumber \\
& \times \int_0^{z_*}\!\!\dd z' \left(\frac{\dd\chi}{\dd z}\right)\,
W_{\nu'}^{(\CIB)}(z')j_{\ell}(k\chi(z')) P^{1/2}_{\rm gg}(k,z',\nu')
\nonumber \,,
\end{align}
where $j_{\ell}(x)$ are spherical Bessel functions.

\subsection{GR corrections and non-Gaussian bias}
\label{sub:GRNG}

There are two important ingredients which we wish to add to the above model: General relativistic 
corrections and non-Gaussian bias. Details about the calculation of the GR effects, which have 
already been considered in the context of intensity mapping by \cite{Hall:2012wd,Alonso:2015uua}, 
can be found in Appendix \S\ref{app:GR}.

At first order in GR perturbations, the observed CIB specific intensity Eq.\ (\ref{eq:CIBNewt})
eventually reduces to
\begin{align}
\label{eq:grCIB}
I(\nu,\nvh) &=
\int_0^{\chi_*}\!\!\dd z\left(\frac{\dd\chi}{\dd z}\right)\, W_{\nu}^{(\CIB)}(z) \\
&\qquad \times \bigg(1 + \delta_\jmath
+ \frac{\partial\ln\bar\jmath_\nu}{\partial\eta}\delta\eta 
+ \delta_\parallel+2\,s\,\delta_\perp\bigg) \nonumber\;,
\end{align}
where $\delta_\jmath$ is the perturbation to the galaxy emissivity in the source rest-frame and 
at constant line-of-sight comoving distance $\chi$, the term proportional to $\delta\eta$ arises 
from the transformation to constant observed redshift, $\delta_\parallel$ is the perturbation to 
the source volume along the line-of-sight, 
and the term proportional to $\delta_\perp$ represents the perturbation to the limiting 
luminosity generated by fluctuations in the luminosity distance. Note that we have introduced a 
magnification bias 
\begin{equation}
\label{eq:magnibias}
s(z) = \frac{\partial\ln\bar\jmath_\nu}{\partial{\bar L}_{_{(1+z)\nu}}^\text{\tiny cut}}
=\frac{{\bar L}_{_{(1+z)\nu}}^\text{\tiny cut}}{4\pi}\frac{\bar n_g}{\bar\jmath_\nu}
\end{equation}
in analogy with source number counts \citep{broadhurst/taylor/peacock:1995}.
The mean comoving emissivity $\bar\jmath_\nu$ is given by Eq.\ (\ref{eq:emod2}).
In the conformal Newtonian gauge adopted for the calculation, the perturbation to the conformal 
time $\eta$ reads
\begin{equation}
{\cal H}\delta\eta= -\Psi-\int_0^\chi\!\!\dd\chi'\,\big(\dot{\Psi}+\dot{\Phi}\big)+\vv\cdot\nvh
\end{equation}
while the perturbations to the source volume element parallel and transverse to the beam are
\begin{align}
\label{eq:dparl}
\delta_\parallel &=\bigg({\cal H}-\frac{\dot{\cal H}}{\cal H}\bigg)\delta\eta+\Psi+\vv\cdot\nvh \\
&\qquad +\frac{1}{\cal H}\bigg[\frac{\dd\Psi}{\dd\chi}+\big(\dot{\Psi}+\dot{\Phi}\big)
-\frac{\dd\vv}{\dd\chi}\cdot\nvh\bigg] \nonumber \;,
\end{align}
and
\begin{equation}
\label{eq:dperp}
\delta_\perp = {\cal H}\delta\eta-\frac{1}{\chi}
\left[\delta\eta-\int_0^\chi\!\!\dd\chi'\,\big(\Psi+\Phi\big)\right] 
-\Phi-\kappa \;,
\end{equation}
respectively.
Here, overdots designate a partial derivative w.r.t.\ the unperturbed conformal time $\eta$. Note 
that $\chi$ and $\eta$ are related through $\chi=\eta_0-\eta$. 
Furthermore,
\begin{equation}
\kappa = \frac{1}{2}\int_0^\chi\!\!\dd\chi'\,\frac{\chi-\chi'}{\chi\chi'}
\nabla_\Omega^2\big(\Psi+\Phi\big) \;,
\end{equation}
where $\nabla_\Omega^2$ is the Laplacian on the unit sphere, is the lensing convergence.
We have restricted ourselves to adiabatic scalar perturbations. There would be more terms, should 
vector or tensor perturbations or isocurvature modes be present at an appreciable level.

Assuming that spatial variations in the comoving emissivity trace fluctuations in the galaxy number
density, intrinsic perturbations to the comoving emissivity $\delta_\jmath$ in the Newtonian gauge
are related to matter fluctuations $\delta_m^\text{\tiny syn}$ in a synchronous gauge comoving with
dark matter through \citep{challinor/lewis:2011,baldauf/seljak/etal:2011,Jeong:2011as}
\begin{equation}
\label{eq:deltanLin}
\delta_\jmath = b_1\, \delta_m^\text{\tiny syn} 
+ \frac{\partial\,\ln\,\bar\jmath_\nu}{\partial\eta} \frac{v}{k} \;,
\end{equation}
where $b_1(k)=b_1^\text{\tiny G}(k)+\Delta b_1^\text{\tiny NG}(k)$ is the linear galaxy bias. 
We have written $b_1(k)$ as generally the sum of a Gaussian piece $b_1^\text{\tiny G}(k)$ (which, in 
the halo model, depends on wavenumber through the profile $u(k,z|M)$) and a scale-dependent correction 
$\Delta b_1^\text{\tiny NG}(k)$ induced by the primordial non-Gaussianity. 
Furthermore, $v$ is a scalar function such that the (curl-free) velocity perturbation  is 
$\vv=k^{-1}\grad v$ in the Newtonian gauge. Finally, we have ignored the 
shot-noise contribution for the reasons stated above. Note that Eq.\ (\ref{eq:deltanLin}) is linear and, 
thus, corresponds to our halo bias $b(M,z)$ in Eq.\ (\ref{eq:Dnu}).

To relate the perturbations $\Psi$, $\Phi$, $\delta_m^\text{\tiny syn}$ and $v$ to the initial conditions, 
we use $\Phi_i(\vk)=(3/5)\zeta(\vk)$, where $\zeta(\vk)$ is the uniform-density gauge curvature perturbation
on super-horizon scales (i.e. $k\ll {\cal H}$), as a reference. 
Our choice follows from the fact that, in the large scale structure literature, the primordial 
non-Gaussianity is commonly laid down immediately after matter-radiation equality 
\citep[see, e.g.,][for a review]{desjacques/seljak:2010}, whence the factor of $3/5$.
The various transfer functions will generally depend on the matter content of the Universe. 
Ignoring anisotropic stresses, these are given by 
\begin{align}
T_\Psi(k,z) &= \bigg. g(z) T(k) \\
T_\Phi(k,z) &= \bigg. T_\Psi(k,z) \nonumber \\
T_\delta(k,z) &= \frac{2}{3\Omega_m}\bigg(\frac{k}{{\cal H}_0}\bigg)^2 T(k) D(z) \nonumber \\
T_v(k,z) &= \bigg. \left(\frac{{\cal H}}{k}\right) f\, T_\delta(k,z) \nonumber \;.
\end{align}
Here and henceforth, $g(z)$ and $D(z)=a(z)g(z)$  are the growth function of potential 
and density perturbations, respectively, $f$ is the logarithmic derivative $f=d\ln D/d\ln a$ and 
the matter transfer function $T(k)$ is obtained with the Boltzmann code {\small CAMB} in a 
synchronous gauge comoving with the pressureless matter.
Note that, with our definitions, the Fourier modes of the peculiar velocity $\vv$ are given by
$\vv(\vk,z)=i\vk {\cal H} f T_\delta(k,z) \Phi_i(\vk)/k^2$.

Non-Gaussianity generated outside the horizon induces a 3-point function that is peaked on squeezed 
or collapsed triangles for realistic values of the scalar spectral index. 
The resulting non-Gaussianity depends only on the local value of the curvature perturbation, and can 
thus be conveniently parameterized by $\Phi_i=\phi+\fnl(\phi^2-\la\phi^2\ra)$,  where $\phi$ denotes 
a Gaussian field \citep{Salopek:1990jq,Gangui:1993tt}. 
The quadratic term introduces a coupling between short- and long-wavelength modes, which results in 
a scale-dependent halo bias at large scales \citep{Dalal:2007cu,Matarrese:2008nc}. 
This non-Gaussian bias takes the form \citep{Slosar:2008hx}
\begin{equation}
\Delta b_1^\text{\tiny NG}(k,z)=
3\fnl \left(\frac{\partial\ln\bar{n}_h}{\partial\ln\sigma_8}\right)
\frac{\Omega_m{\cal H}_0^2}{a(z)\, T_\Phi(k,z) k^2}\;.
\label{eq:bshift}
\end{equation}
It is this bias enhancement on large scales that we are trying to measure with the CIB. A large, 
potentially detectable $\fnl\gtrsim 1$ can be produced e.g. by multiple scalar fields
\citep{linde/mukhanov:1997,lyth/ungarelli/wands:2003}. 

We have assumed that the halo mass function is universal and, thus, replaced the logarithmic derivative 
of the mass function by $\dc(b_1(M,z)-1)$, where $\dc \sim 1.68$ is the present-day (linear) critical 
density threshold. For the linear Gaussian bias, we use the fitting formula given in \cite{tin10a}.

Departure from Statistical Gaussianity in the initial conditions significantly affect the abundance of
highly biased tracers of the LSS, since their frequency sensitively depends on the tail of the density
PDF \citep[e.g.][]{lucchin/matarrese:1988}. To ascertain the importance of this effect, we have replaced
the halo mass function in Eq.\ (\ref{eq:meanj}) by
\begin{equation}
\bar n_h(M,z,\fnl)=\bar n_h^\text{\tiny G}(M,z) \Big(1+R(M,z,\fnl)\Big) \;,
\end{equation}
where $\bar n_h^\text{\tiny G}(M,z)$ is our fiducial Gaussian, Tinker mass function and $R(M,z,\fnl)$ is 
the non-Gaussian fractional correction modelled along the extensions proposed by 
\cite{matarrese/verde/jimenez:2000,loverde/miller/etal:2008} (see \cite{desjacques/seljak:2010} for a
discussion). We have found that, for an input value $\fnl=1$, including $R(M,z,\fnl)$ amounts to a 
$\sim 1$\% correction to the amplitude of the non-Gaussian CIB bias. We will thus ignore this effect in
what follows.

\subsection{Signature of PNG in the CIB anisotropies}
\label{ssec:png}

\begin{figure*}
\centering
\includegraphics[width=80mm]{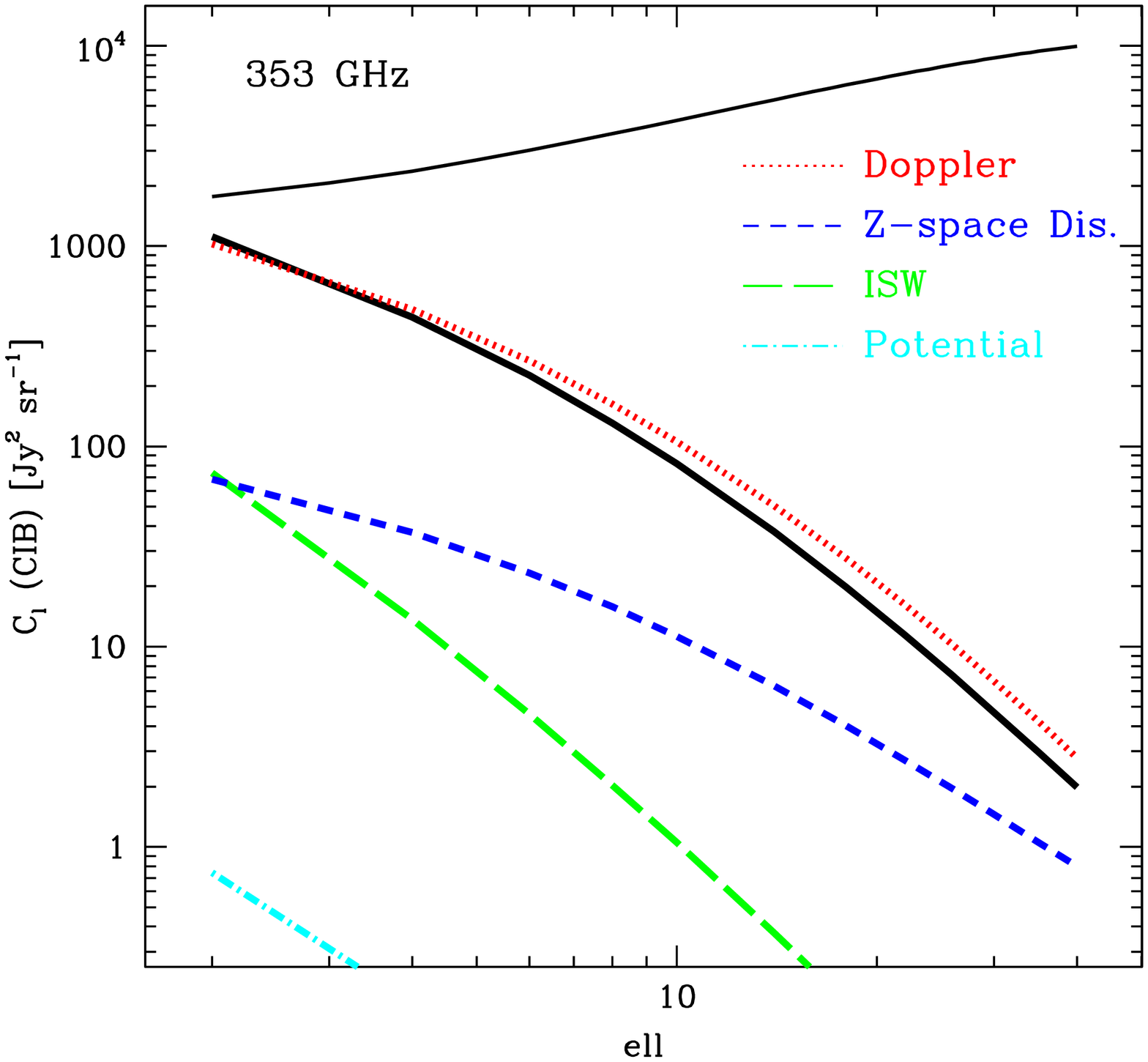}
\includegraphics[width=80mm]{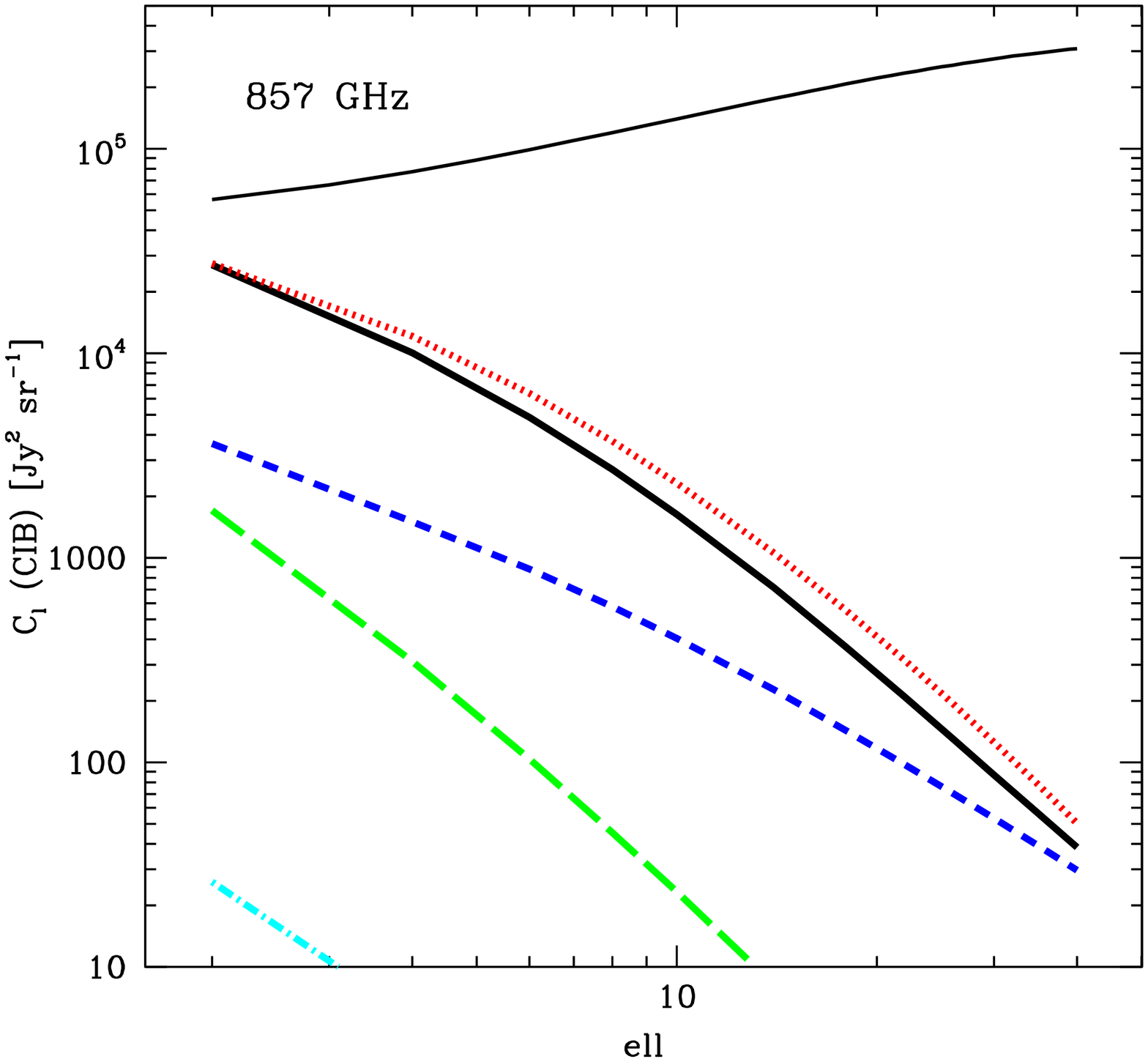}
\caption{GR corrections (solid thick lines) to the CIB power spectrum
  (solid thin lines) at 353 (left panel) and 857\,GHz (right
  panel). The different GR contributions are also isolated: the
  velocity term (dotted red lines); the redshift--space distorsion
  (short dashed blue lines); the ISW term (long dashed green term);
  the potential term (dot--dashed cyan lines). The GR corrections are
  relevant at the very first multipoles, and become lower than 1\,per
  cent at $\ell>10$. The dominant contribution comes from the velocity
  term.}
\label{fig:cl1}
\end{figure*}

\begin{figure*}
\centering
\includegraphics[width=80mm]{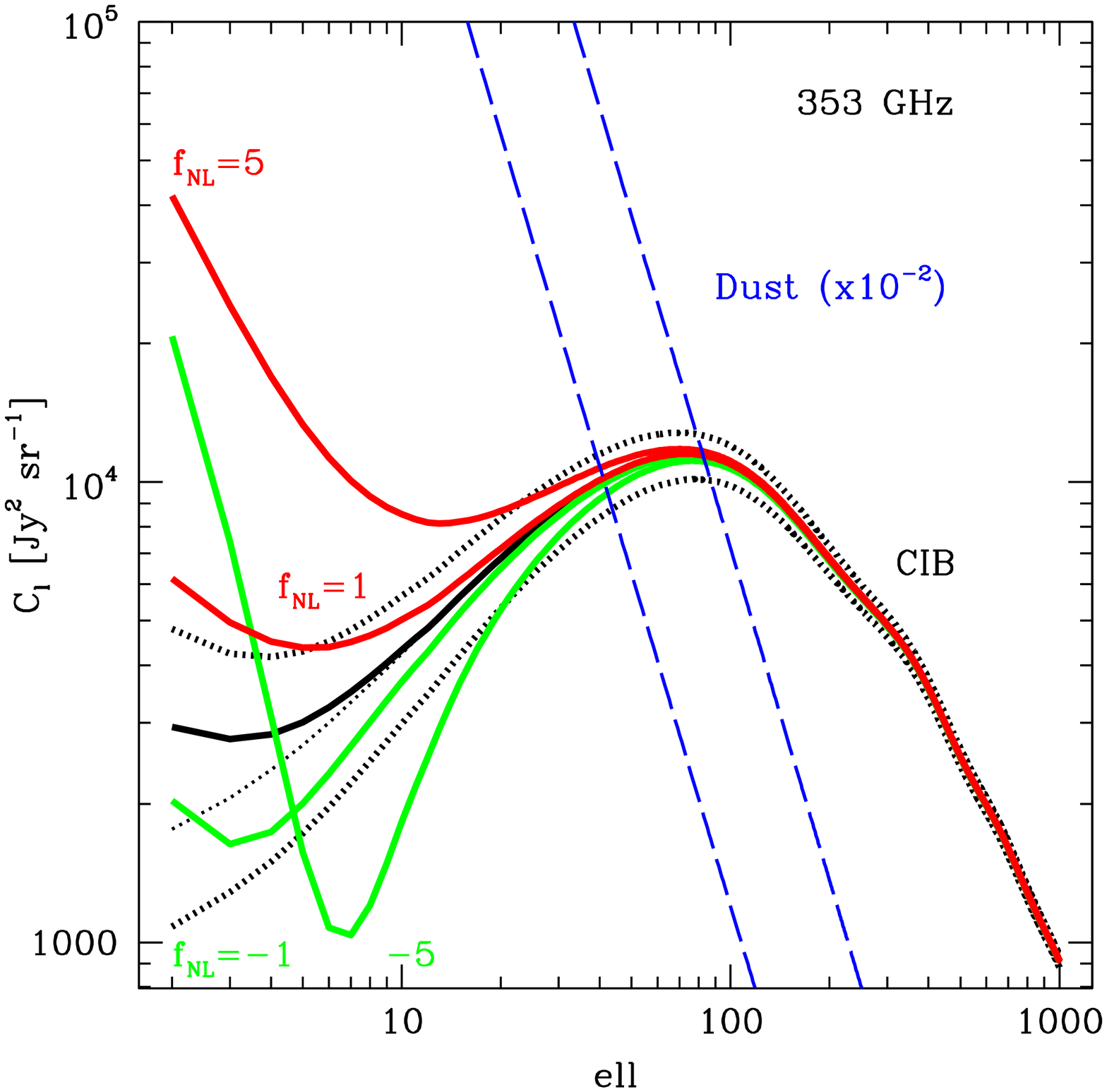}
\includegraphics[width=80mm]{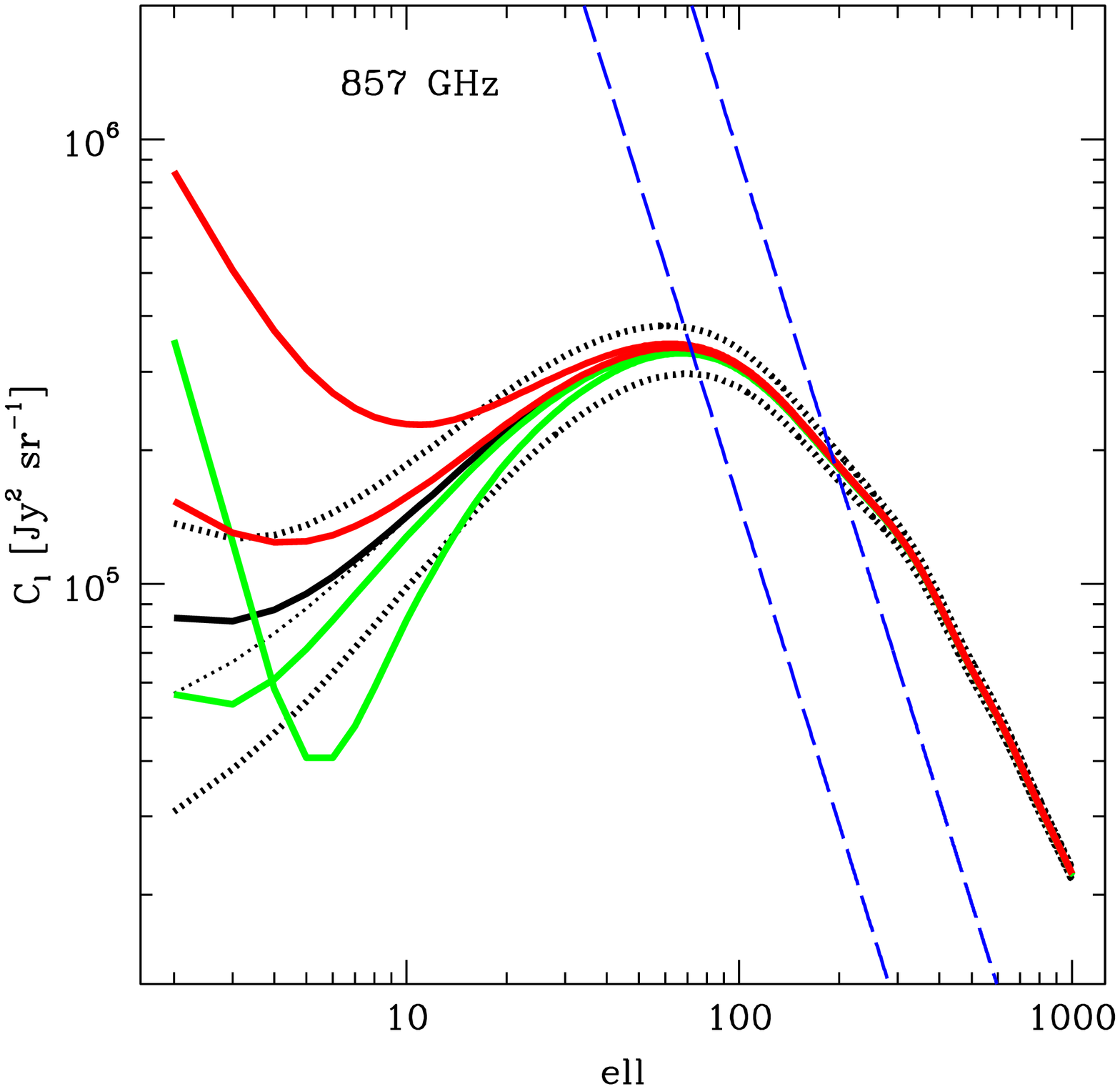}
\caption{CIB power spectra (including GR corrections) at 353 and
  857\,GHz assuming Gaussian primordial fluctuations (black solid
  lines) and $\fnl=\pm1$ and $\pm5$ (solid red and green lines). 
  The two, upper and lower thick dotted lines indicate the cosmic 
  variance associated to the Gaussian case, whereas the median, 
  thin dotted curve is the CIB power spectrum without the GR 
  corrections. Also shown as blue dashed line is the dust contamination -- 
  reduced by a factor 100 -- as measured by \citet{planck_xxii_2015} for the
  cleanest sky patches with coverage fraction 10 (left line) and 
  40\,per\,cent (right line).
  The dust contamination is still orders of magnitude larger than
  CIB fluctuations on large angular scales.
  At $\ell=2$ the CIB spectrum increases by about an order of 
  magnitude as we turn on $\fnl$ from 0 to $\pm5$. 
  At larger multipoles, the effect of non--Gaussianity quickly decreases, 
  and it is lower than cosmic variance at $\ell\gtrsim 20$ even for 
  $\fnl\sim5$. 
  For negative values of $\fnl$, the CIB power spectrum presents a minimum 
  at $\ell\lsim10$ corresponding to the angular scale at which the NG 
  bias contribution becomes dominant over the Gaussian bias term.}
\label{fig:cl2}
\end{figure*}

\begin{figure}
\centering
\includegraphics[width=80mm]{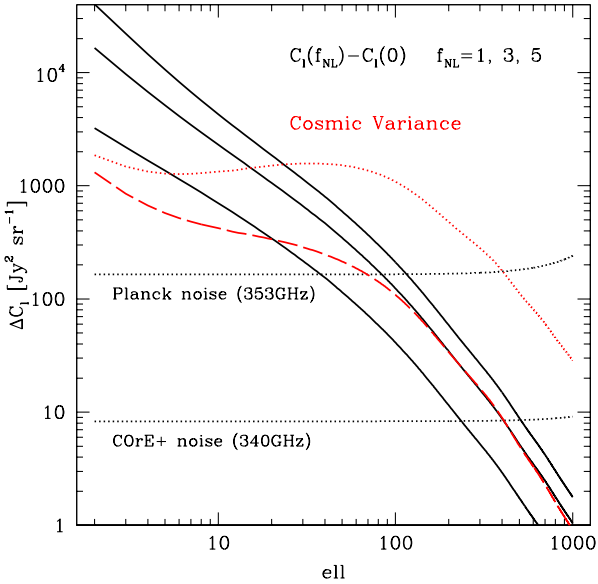}
\caption{The cosmic variance associated to the CIB power spectrum with
  Gaussian initial conditions at 353\,GHz (red dotted line) is shown in
  comparison with the PNG signal $\Delta
  C_{\ell}=C_{\ell}(\fnl)-C_{\ell}(0)$ for $\fnl=1$, 3 and 5 (black
  solid lines). The red dashed line gives the cosmic variance reduced
  by a factor $\sqrt{\ell}$, corresponding to power spectra binned in
  multipole intervals of size $\ell$. The cosmic variance becomes
  lower than $\Delta C_{\ell}$ at $\ell\lsim20$ for $\fnl=1$ and
  lower or equal to $\Delta C_{\ell}$ up to $\ell\ga1000$ for
  $\fnl=3$. We also plot the noise level for the \planck\ and
  COrE+ channels at $\nu\sim350\,$GHz (black dotted lines).}
\label{fig:cl3}
\end{figure}

\begin{figure}
\centering
\includegraphics[width=80mm]{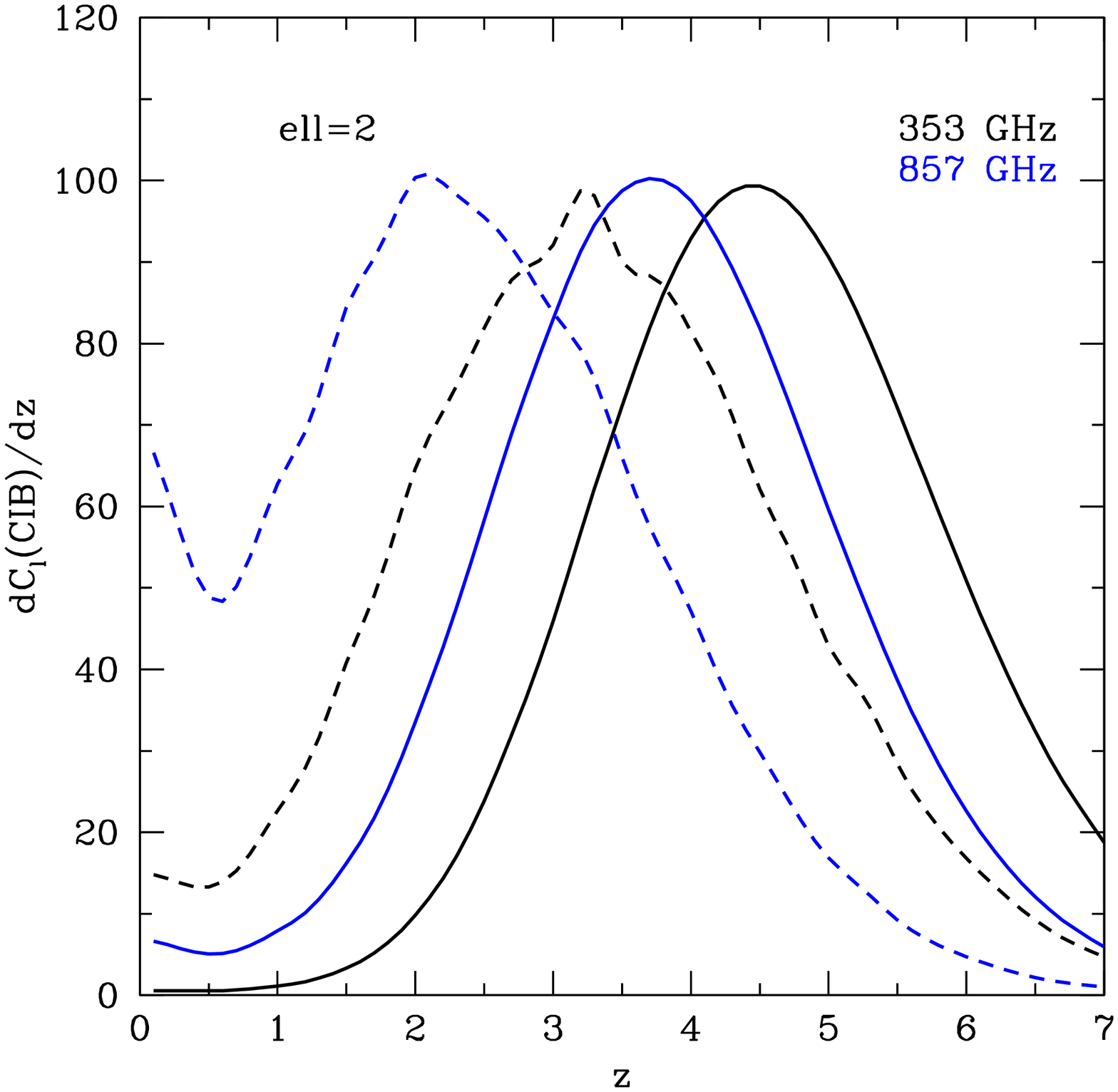}
\caption{The redshift kernel for the CIB power spectrum at $\ell=2$ for
  two different frequencies: 353\,GHz (black lines) and 857\,GHz (blue 
  lines). Dashed lines are for Gaussian primordial fluctuations, whereas 
  solid lines assume local PNG with $\fnl=5$. CIB anisotropies induced 
  by PNG mainly arise from dusty galaxies at redshifts between 
  $2.5\lsim z\lsim6$ with a mild dependence on frequency.}
\label{fig:cl4}
\end{figure}

The signature imprinted by the scale-dependent, non-Gaussian halo bias 
is largest at low multipoles where the contribution of the 1-halo term
is negligible. We have indeed checked that, for the frequencies 
considered in our forecast, the 1-halo term is at most $\lesssim 10$\%
of the 2-halo contribution. This corresponds to an effective nonlinear 
parameter $|\fnl|\lesssim 0.1$, which is hardly accessible to CMB 
experiments with the sensitivity considered here.
Therefore, we can safely neglect it at the low multipoles $\ell\leq 40$ 
where Eq.\ (\ref{eq:emod6}) is applied. Notwithstanding, we include it at 
the higher multipoles used to constrain the CIB model parameters since 
it dominates over the 2-halo term for $\ell\gtrsim 1000$. 

At low multipoles $\ell\lesssim 40$ where the Limber approximation is
not valid, we ignore the 1-halo term, but retain the GR corrections 
and the non-Gaussian bias so that the CIB angular power spectrum is 
computed as
\begin{align}
\label{eq:cl}
C^{(\CIB)}_{\ell}(&\nu,\nu') = \frac{2}{\pi}\int\!\! \dd k\,k^2 \\
&\times \int_0^{z_*}\!\!\dd z\left(\frac{\dd\chi}{\dd z}\right)\,
W_{\nu}^{(\CIB)}\!(z)F_\ell(\nu,k,\chi(z))P^{1/2}_{\Phi_i}\!(k) 
\nonumber \\
& \times \int_0^{z_*}\!\!\dd z' \left(\frac{\dd\chi}{\dd z'}\right)\,
W_{\nu'}^{(\CIB)}\!(z')F_\ell(\nu,k,\chi(z'))P^{1/2}_{\Phi_i}\!(k)
\nonumber \;.
\end{align}
Here, $P_{\Phi_i}\!(k)$ is the power spectrum of our reference curvature
pertubation, and is related to the linear matter power spectrum through
\begin{equation}
P_\text{lin}(k,z) = \frac{4}{9\Omega_m^2}\left(\frac{k}{{\cal H}_0}\right)^4
T^2(k) D^2(z) \, P_{\Phi_i}\!(k) \;.
\end{equation}
%
Furthermore, the radial window function $F_\ell(\nu,k,\chi)$ is given by
\begin{align}
F_\ell(&\nu,k,z) = j_\ell(k\chi)\bigg[b_1(\nu,k,z) T_\delta
+\frac{\partial\ln\bar\jmath_\nu}{\partial\eta}
\left(\frac{T_v}{k}-\frac{T_\Psi}{\cal H}\right) 
+\frac{\dot{\cal H}}{{\cal H}^2}T_\Psi
\nonumber \\
&+\frac{1}{\cal H}\Big(\dot{T}_\Psi+\dot{T}_\Phi\Big)\bigg] 
+ j_\ell'(k\chi)\bigg[\left(\frac{1}{\cal H}\frac{\partial\ln\bar\jmath_\nu}{\partial\eta}
-\frac{\dot{\cal H}}{{\cal H}^2}+2\right) T_v
\nonumber \\
&+\frac{k}{\cal H}T_\Psi\bigg]- j_\ell''(k\chi)\left(\frac{k}{\cal H}\right)T_v 
-\bigg(\frac{1}{\cal H}\frac{\partial\ln\bar\jmath_\nu}{\partial\eta}-\frac{\dot{\cal H}}{{\cal H}^2}+1\bigg) 
\nonumber \\
&\quad \times \int_0^\chi\!\!\dd\chi'\,j_\ell(k\chi')\big(\dot{T}_\Psi+\dot{T}_\Phi\big) \;,
\label{eq:Fl}
\end{align}
with the linear bias 
\begin{align}
b_1(&\nu,k,z) = \int\!\!\dd M\,
\bigg[b_1^\text{\tiny G}(M,z)+3\fnl\dc\big(b_1^\text{\tiny G}(M,z)-1\big)
\frac{\Omega_m{\cal H}_0^2}{a\, T_\Phi\,k^2}\bigg] \nonumber \\
&\qquad \times \bar n_h(M,z) \Bigl[f_\nu^\text{c}(M,z)+f_\nu^\text{s}(M,z)\Bigr]\,
u(k,z|M)
\end{align}
predicted by the halo model and including the correction due to primordial 
non-Gaussianity. Details about the calculation of $F_\ell(\nu,k,z)$ can be found 
in Appendix \S\ref{app:cl}. 
Note that Eq.\ (\ref{eq:Fl}) does not include the lensing magnification induced by 
the luminosity cut--off, see Eq.\ (\ref{eq:emod2}).
The typical flux density cuts in CMB maps indeed are of the order of 
$\approx100$\,mJy, which correspond to luminosities $>10^{27}$W\,Hz$^{-1}$ 
at $z>1$. 
These luminosities are much larger than the expected luminosity of submillimeter 
galaxies.

In Fig.\,\ref{fig:cl1}, the GR corrections to the CIB power spectrum are shown 
at frequency 353 and 857\,GHz. 
As expected, these corrections are relevant only at the very large angular 
scales, $\ell\lsim10$. At $\ell=2$, they are of the same order of magnitude as
the CIB spectrum (assuming Gaussian primordial fluctuations) and
decrease to 1--2\,per cent at $\ell=10$, with little dependence on the
frequency. GR corrections are dominated by the velocity term, whereas
the other terms can be considered negligible\footnote{In
  Figure\,\ref{fig:cl1}, the contribution to GR corrections from a
  single term are computed neglecting the other components. It should
  be noted that the total GR correction is not simply the sum of the
  single contributions because the double products among different
  terms in Eq.\ (\ref{eq:cl}) have to be included.}.

The effect of primordial non--Gaussianity on the CIB power spectrum is much 
more pronounced than the imprint of GR corrections so long as $|\fNL|\bsim 1$. 
For $|\fnl|\lesssim 1$, the GR corrections must be included in the analysis 
to avoid significant bias on $\fnl$ (see \cite{Camera:2014sba} for related
discussion in the context of galaxy redshift surveys).
In Figure\,\ref{fig:cl2}, we compare the CIB spectrum assuming $\fnl=\pm5$, 
$\pm1$ and 0, i.e. with primordial Gaussian conditions. GR corrections are 
also included (see the small increment in CIB spectra with $\fnl=0$ at the 
first $\ell$s). Like GR corrections, the non-Gaussian bias leads to a 
pronounced signal at low multipoles ($\ell\lsim10$), which decreases 
proportionally to $\ell^{-1}$. Namely, a local PNG with $\fnl\sim5$ increases
the CIB power spectrum by one order of magnitude at $\ell=2$, and by a factor
of $\lsim2$ at $\ell=10$. When $\fnl<0$, the CIB spectrum has typically less
power than in the Gaussian case. It reaches a minimum at some multipole (due 
to a cancellation in the 2-halo term) before increasing again at very low 
$\ell$. This leads to a characteristic feature in the CIB power spectrum.

To gain some insight into the redshift and frequency dependence of the 
non-Gaussian signal, let us assume that the luminosity-mass relation can
be approximated by a Dirac delta, i.e. 
$\Sigma(M)\approx \delta_D(M-M_\text{eff})$. In this case, the non-Gaussian 
CIB bias $\Delta b_1^\text{\tiny NG}(\nu,k,z)$ simplifies to
\begin{align}
\label{eq:CIBNG}
\Delta b_1^\text{\tiny NG}(\nu,k,&z) \approx
\frac{3}{4\pi} \fnl \frac{\Omega_m{\cal H}_0^2}{a\, T_\Phi\,k^2}
L_0 (1+z)^\delta \Theta_\CIB\big[(1+z)\nu\big] 
\nonumber \\
&\quad \times \frac{\partial}{\partial\ln\sigma_8}\bigg[
\bar n_h(M_\text{eff},z)\Theta_H(M_\text{eff}-M_\text{cen}) 
\nonumber \\
&\qquad + \int_{M_\text{eff}}^\infty\!\!dM \bar n_h(M,z)
\bar n_s(M_\text{eff},z|M)\bigg] \;,
\end{align}
where $\Theta_H$ is the Heaviside step function.
The integrand of the second term in the square bracket will be maximized
for some halo mass $M=\alpha M_\text{eff}$ with $\alpha > 1$ and, thus,
can be approximated as 
$\approx \bar n_h(\alpha M_\text{eff},z) \bar n_s(M_\text{eff},z|\alpha M_\text{eff})$.
Note that $\alpha$ will generally depend on redshift. 
A similar calculation yields 
\begin{align}
b_1^\text{\tiny G}(\nu,k,z) &\approx
\frac{L_0}{4\pi}(1+z)^\delta \Theta_\CIB\big[(1+z)\nu\big]
\\
&\quad \times  
\bigg[\bar n_h(M_\text{eff},z)b_1^\text{\tiny G}(M_\text{eff},z) 
\Theta_H(M_\text{eff}-M_\text{cen}) 
\nonumber \\
&\qquad +
\int_{M_\text{eff}}^\infty\!\!dM b_1^\text{\tiny G}(M,z) \bar n_h(M,z)
\bar n_s(M_\text{eff},z|M)\bigg] \nonumber
\end{align}
for the Gaussian part of the CIB bias. Hence, ignoring the contribution from 
the satellite galaxies, the relative amplitude of the non-Gaussian CIB bias 
scales like
\begin{align}
\frac{\Delta b_1^\text{\tiny NG}}{b_1^\text{\tiny G}}(\nu,k,z) &\sim 
3\fnl \frac{\Omega_m{\cal H}_0^2}{a\, T_\Phi\,k^2} 
\big(b_1^\text{\tiny G}(M_\text{eff},z)\big)^{-1} \\
&\qquad \times
\frac{\partial\ln\bar n_h}{\partial\ln\sigma_8}(M_\text{eff},z)\,
\Theta_H(M_\text{eff}-M_\text{cen}) \nonumber \;,
\end{align}
i.e. it does not depend on $\nu$. Therefore, the relative amplitude of the signal 
will weakly depend on the value of $\delta$ or the exact shape of 
$\Theta_\CIB$ which, in our model, do not depend on halo mass. 
We thus expect that the CIB non-Gaussian bias mainly depends on the HOD parameters, 
i.e. $M_\text{eff}$, $M_\text{cen}$ and the distribution of subhalos hosting 
galaxies. We will ascertain the sensitivity to HOD modeling in more
detail in \S\ref{sec:discussion}.

\subsection{Galactic dust and cosmic variance}

Figure\,\ref{fig:cl2} shows the two major constraints for the
detection of PNG in CIB spectra: the Galactic dust emission and the
cosmic variance. The former dominates the CIB emission at all the
relevant frequencies by orders of magnitude, especially on the largest
angular scales. Its power spectrum is typically $\propto\ell^{-2.4}$
and the amplitude is strongly dependent on the area of the sky
considered. 
We will extensively discuss this point later in
\S\ref{sec:dust}. The second main constraint comes from the cosmic
variance. We can note that $|\fnl|\bsim1$ is required in order to have
corrections to CIB spectra larger than the cosmic variance associated
to the ``standard'' CIB spectrum. The signal is, in any case,
tiny already at $\ell\ga20$ even for larger $\fnl$ and typically below
the cosmic variance. However, cosmic variance can be significantly
reduced by combining information from independent multipoles. As
example, if we assume to measure the CIB power spectra in bins of
width $\ell$, the cosmic variance will decrease by a factor
$\sqrt{\ell}$, as shown in Figure\,\ref{fig:cl3}. The cosmic variance
will be then of the same level as or lower than the PNG signal for
$|\fnl|\ga3$ at all the multipoles considered.

It is well known that the redshit distribution of dusty galaxies which
mainly contribute to CIB anisotropies slowly changes with frequency.
Namely, it moves to higher redshifts when the CIB is observed at
longer wavelengths \citep[e.g.,][]{planck_xviii_2011}. For example, we
show in Figure\,\ref{fig:cl4} the redshift integrand for the CIB
spectra at $\ell=2$: the peak shifts from redshift 2 to $\sim3$ when
the frequency changes from 857 to 353\,GHz. In both cases, however,
the contribution from high--redshift galaxies ($z\bsim4$) is not
negligible. It is interesting to consider the same in the presence of
PNG (in Figure\,\ref{fig:cl4}, $\fnl=5$). At $\ell=2$, where the CIB
spectrum is dominated by anisotropies induced by PNG, the integrand
peaks even at higher redshifts, i.e. at $z=3.5$--4.5 according to the
frequency. CIB anisotropies induced by PNG are therefore mainly from
dusty galaxies at redshifts significantly higher than for the
``standard'' CIB, covering a redshift interval between $2\lsim
z\lsim7$. 

\section{Fisher matrix formalism}
\label{sec:fish}


We adopt a Fisher matrix approach to provide estimates of the
sensitivity of CIB measurements to primordial non--Gaussianity. In
terms of power spectra, the Fisher matrix element $F_{ij}$ can be
written as \citep[e.g.,][]{teg00}
\begin{equation}
F_{ij}=\sum_{\ell=\lmin}^{\lmax}\,\frac{2\ell+1}{2}f_\text{sky}\,
{\rm Tr}\Bigg({\bf C}_{\ell}^{-1}\frac{\partial{\bf C}_{\ell}}{\partial\theta_i}
{\bf C}_{\ell}^{-1}\frac{\partial{\bf C}_{\ell}}{\partial\theta_j} \Bigg)\,,
\label{eq:fisher}
\end{equation}
The model parameters $\theta_i$ and $\theta_j$ include the CIB model 
parameters given in Table\,\ref{tab:0} plus the primordial 
non--Gaussianity parameter $\fnl$ and, when considered, the parameters 
related to the dust emission. 
The covariance matrix $\bf{C}_{\ell}$ is an $N_{\nu}\times N_{\nu}$ 
matrix whose elements are defined as the auto-- and cross--power spectra 
of data at $N_{\nu}$ different observational
frequencies, i.e. $(\bf{C}_{\ell})_{ij}=C_{\ell}^{\nu_i\nu_j}$. In
absence of residual foregrounds and CMB radiation,
$C_{\ell}^{\nu_i\nu_j}=C_{\ell}^{(\CIB)}(\nu_i,\nu_j)+N_{\ell}(\nu_i)\delta_{ij}$,
where the diagonal elements of the covariance matrix contain the
Gaussian instrumental noise terms:
\begin{equation}
N_{\ell}(\nu)=w^{-1}_{\nu}\exp{\Bigg(\ell(\ell+1)
\frac{\theta^2_{\FWHM}(\nu)}{8\log2}\Bigg)}\,,
\label{eq:noise}
\end{equation}
where $w^{-1/2}_{\nu}$ is the instrumental white noise level in
Jy\,sr$^{-1/2}$ and $\theta_{\FWHM}$ is the full-width at
half-maximum beam size in radians at the frequency $\nu$. In
Eq.\ (\ref{eq:fisher}), $f_\text{sky}$ is the fraction of the sky used
to recover CIB fluctuations, and the factor $(2\ell+1) f_\text{sky}$
gives the effective number of uncorrelated modes per multipole.

The uncertainty on $\fnl$ is computed after marginalizing over the
other parameters, i.e. upon inverting the Fisher matrix so that
$\sigma(\fnl)=\sqrt{[F^{-1}]_{11}}$ (where $i=1$ in the Fisher matrix
corresponds to the $\fnl$ parameter). In Eq.\,(\ref{eq:fisher}), we
define the smallest observable multipole for an experiment to be
$\lmin=\pi/(2\fsky^{1/2})$, rounding up to the next integer. As
maximum multipole, we assume $\lmax=1000$. This guarantees that
the shot--noise contribution from star--forming dusty galaxies is
negligible. Moreover, multipoles $\ell\ga1000$ provide negligible
information on $\fnl$.

As a first step, we apply the Fisher formalism to the ideal case of
CIB maps from which (Galactic) foregrounds and CMB have been perfectly
removed. We refer to the instrumental configuration of a possible
future CMB mission like
COrE+\footnote{http://conservancy.umn.edu/handle/11299/169642}, as
reported in Table\,\ref{tab:3}. Hereafter, only frequencies higher
than 200\,GHz will be considered in the analysis. Channels at lower
frequencies are in fact dominated by the CMB and should be dedicated
to removing CMB fluctuations from the signal.

This preliminarly test gives us an idea of the maximum level at which 
$\fnl$ can be detected through CIB observations by future space missions 
dedicated to the measurement of the CMB polarization. 
In Table\,\ref{tab:1} we
report the uncertainty on $\fnl$ assuming $\fnl=0$, using 40\% of the
sky and a different number of frequencies. $\sigma(\fnl)$ is weakly
dependent on the value of $\fnl$. We see that, in principle, $|\fnl|$
of 1--2 could be detectable with high significance ($\ga3$--$\sigma$)
using 2--4 frequency channels, while $|\fnl|$ lower than 1 is
accessible only if $N_{\nu}>4$. We also quote the results obtained 
after neglecting  the first 20 multipoles, which are the most affected 
by Galactic dust residuals. While we observe a strong degradation in the
$\fnl$ sensitivity when only 1 or 2 frequency channels are employed, the
uncertainty on $\fnl$ only marginally increases for $N_{\nu}\ge4$. This
proves that the very large angular scales are not strictly required to 
detect the PNG signal.

We emphasize that an estimate of CIB fluctuations at different frequencies 
is crucial to measure $\fnl$. Apart from giving a better control of the 
foregrounds, it allows 1) a reduction of the noise level by a factor 
$\approx\sqrt{N_{\nu}}$ (if all channels have the same sensitivity) and 2) 
an improvement of the determination of CIB parameters 
(with $\sigma(\fnl)\rightarrow1/F_{11}$ for $N_{\nu}>2$). 
In addition, because CIB anisotropies are not perfectly correlated at 
different frequencies, multi--frequency CIB observations allow to combine 
signals from partly overlapping volumes of the Universe. In the overlapping
regions, these measurements trace similar matter fluctuations. Therefore,
they can be combined to decrease cosmic variance and improve the 
signal-to-noise ratio, analogously to the multi-tracer technique in galaxy
clustering 
\citep[see e.g.][for applications to $\fnl$]{seljak:2009,Hamaus:2011dq}.

\begin{table}
\begin{center}
  \caption{Uncertainty on $\fnl$ obtained from a COrE+--like
    experiment in purely CIB+noise maps, using 40\% of the sky, with
    different frequency combinations and $\lmin=3$ and
    20. \label{tab:1}}
\begin{tabular}{ccccc}
\hline
& & & \multicolumn{2}{c}{$\sigma(\fnl=0)$} \\
\hline
\# & $\nu$\,[GHz] & & $\lmin=3$ & $\lmin=20$ \\
\hline
1 & 340 & & 0.68 & 3.0 \\
2 & 340,520 & & 0.52 & 1.4 \\
4 & $[220,520]$ & & 0.36 & 0.50 \\
8 & $[220,600]$ & & 0.21 & 0.22 \\
\hline
\end{tabular}
\end{center}
\end{table}

\section{Forecasts including Galactic dust contamination}
\label{sec:dust}

\begin{figure}
\centering
\includegraphics[width=80mm]{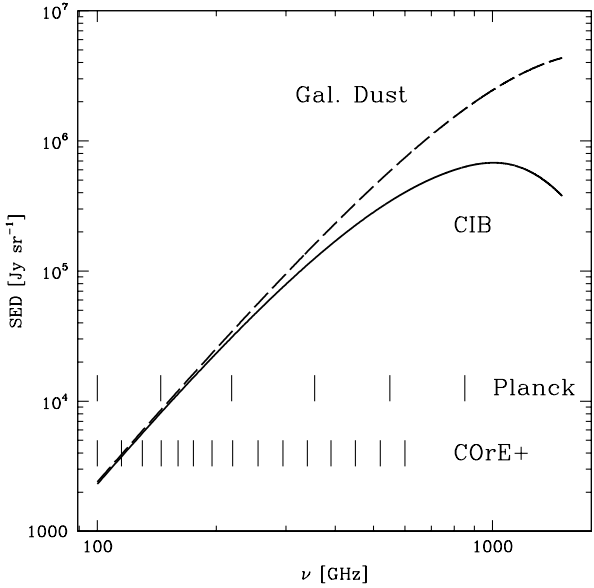}
\caption{Frequency spectrum of the CIB brightness (solid line) and of
  the Galactic dust emission (dashed line) according to the models
  used in the paper. Dust spectrum is normalized to the CIB brightness
  at 100\,GHz. The observational frequencies of \planck\ and COrE+ are
  also shown.}
\label{fig:sed}
\end{figure}

Measuring CIB anisotropies is a challenging task, especially over 
large areas of the sky, even for future high--sensitivity CMB space 
missions. CIB fluctuations are a sub--dominant component at all 
frequencies: CMB anisotropies dominate at frequencies $\nu<200\,$GHz 
(and over the CIB up to $\lsim350$\,GHz), whereas Galactic dust 
emission is preponderant at higher frequencies. 
On the one hand, low--frequency templates are quite effective at 
subtracting the CMB component from maps at few hundreds of GHz 
\citep[see, e.g.][for a detailed discussion about this point]{planck_xxx_2013}. 
On the other hand, distinguishing Galactic from extragalactic dust 
emission is more difficult because of their fairly similar spectral 
energy distribution (SED) which approximately scales in both cases 
like a modified blackbody law. 
In Fig.\,\ref{fig:sed}, we plot the SED for the CIB intensity compared
to the Galactic dust frequency spectrum (see below for the model
description), normalized to the 100--GHz CIB total intensity. The
spectra are almost identical at frequencies below $300$\,GHz, but
the CIB spectrum flattens at higher frequencies, peaking at
about 1000\,GHz. 
Therefore, the two SEDs differ most in the Far--Infrared frequencies, 
suggesting that observations at $\gg500$\,GHz will be crucial to 
separate these two components.

On large areas of the sky, Galactic dust is orders of magnitude brighter
than the CIB. Furthermore, it largely dominates the CIB at multipoles 
$\ell\lsim200$ in the cleanest regions of the sky (see Figure\,\ref{fig:cl2}).
Extracting the CIB signal from Galactic contamination thus requires a 
very accurate component separation. Currently, typical component 
separation methods rely on N$_{\rm HI}$ maps, which furnish a tracer 
of the dust gas \citep{planck_xxx_2013,planck_xviii_2014}. 
However, this method has only been applied successfully over limited 
areas of the sky, leaving always Galactic dust residuals at the level 
of 5--10\%. These residuals strongly affects angular scales $\ell<200$. 
Component separation methods that exploit not only the frequency spectral 
information but also the spatial information are now being developed, 
with the aim to separate the two components over large areas of the sky 
in the \planck\ data \citep{planck_cib_2016}.

\subsection{Model of the Galactic thermal dust emission}
\label{ssec:mod}

Based on the latest \planck\ results \citep{planck_x_2015}, the
frequency spectrum of the Galactic thermal dust can be accurately
described by a modified blackbody
\begin{equation}
\Theta_d(\nu)=\bigg(\frac{\nu}{\nu_0}\bigg)^{\beta_d+3}
\frac{\exp(\gamma\nu_0)-1}{\exp(\gamma\nu)-1}\,,
\label{eq:mod1}
\end{equation}
where $\nu_0$ is the reference frequency (we choose $\nu_0=353$\,GHz,
if not otherwise specified) and $\gamma=2\pi \hbar/k_B T_d$.
The emissivity index
$\beta_d$ and the dust temperature $T_d$ can vary on a pixel-by-pixel
basis depending on the dust population and environment. 
We fix them to the mean values found by \planck\ (\citealt{planck_xxii_2015};
\citealt{planck_x_2015}), $\beta_d=1.53$ and $T_d=19.6$\,K.

The intensity of the dust emission strongly depends on the area of the
sky considered, with high--latitude regions being noticeably less affected
vy dust contamination. In \citet{planck_xxii_2015}, the 
dust angular power spectrum at 353\,GHz has been measured at low
multipoles, $\ell<100$, as a function of the Galactic mask (with
$\fsky$ varying from 40 to 80 per cent). The resulting power spectrum is 
well fitted by a power law with a slope consistent with $-2.4$ over all 
the Galactic masks. However, the amplitude varies nonlinearly as a function 
of $\fsky$. Following their Eq.\,(D.3), for $\fsky$ between 0.4 and 0.8, 
we model the 353--GHz dust power spectrum (in unit of Jy$^2$\,sr$^{-1}$) as
\begin{eqnarray}
C^d_{\ell}(353) & = & A_d\,\Bigg(\frac{\ell}{100}\Bigg)^{\alpha_d}
\nonumber \\
A_d (\fsky) & = & 1.45\times10^6\,
\Bigg(\frac{\fsky}{0.6}\Bigg)^{[4.60+7.11\ln(\fsky/0.6)]}\,,
\label{eq:mod2}
\end{eqnarray}
with $\alpha_d=-2.4$. The amplitude of the spectrum decreases
therefore by a factor $\sim7$ from $\fsky=0.8$ to 0.6, and by a
factor 2 from 0.6 to 0.4 (see Table\,\ref{tab:2}).

Moving to smaller areas of the sky, we can get an estimate of the
amplitude of dust power spectra from the empirical relation
$C^d_{\ell}(\nu)\propto\langle I_{\nu}\rangle^2$, where $\langle
I_{\nu}\rangle$ is the average dust intensity in the sky region considered.
This relation, which is somewhat expected, has been tested with data 
at different frequencies \citep{miv07,planck_xxx_2015}. 
\citet{planck_xxx_2015} estimated $\langle I_{\nu}\rangle$ over large 
regions of the sky with $\fsky$ from 0.24 to 0.72, and over 352 patches 
of 400\,deg$^2$ ($\fsky\simeq0.01$) at Galactic latitude $|b|>35^{\circ}$. 
The dust mean intensity was found to lower from 0.167 to 0.106\,MJy\,sr$^{-1}$
as $\fsky$ varies from 0.63 to 0.42. 
Based on the previous scaling relation, this translates into a decrease 
in the power spectrum amplitude of a factor 2.5, very close to the value 
given by Eq.\,(\ref{eq:mod2}) (i.e., 2.65). 
In the region with $\fsky=0.24$, the measured mean intensity is 
0.068\,MJy\,sr$^{-1}$. This value is only about a factor 1.5--2 higher 
than the mean intensity in the cleanest 400\,deg$^2$ patches 
(0.04--0.045\,MJy\,sr$^{-1}$). 
Using the measured $\langle I_{\nu}\rangle$, we estimate $A_d$ for $\fsky<0.4$, 
as reported in Table\,\ref{tab:2}. In particular, the value of $A_d$ for 
$\fsky=0.1$ has been obtained assuming a sky fraction of 10\% and a dust
contamination as low as the cleanest 400\,deg$^2$ patches observed by
\planck. This is an indicative, albeit maybe optimistic, estimate.
\begin{table}
\begin{center}
  \caption{Estimated amplitude of the dust angular power spectrum at
    353\,GHz as a function of the fraction of the sky. \label{tab:2}}
\begin{tabular}{cccccc}
\hline
$\fsky$ & 0.8 & 0.6 & 0.4 & 0.2 & 0.1 \\
\hline 
$A_d/10^6$ [Jy$^2$\,sr$^{-1}$] & 9.78 & 1.45 & 0.72 & 0.29 & 0.12 \\
\hline
\end{tabular}
\end{center}
\end{table}

It is convenient to factor the auto-- and cross-- dust power spectra
into a spatial term (Eq.\,\ref{eq:mod2}), a frequency--dependence term
(Eq.\,\ref{eq:mod1}), and a frequency--correlation term:
\begin{equation}
C^d_{\ell}(\nu,\nu')=C^d_{\ell}(\nu_0)\,\Theta_d(\nu)\Theta_d(\nu')\,
R_{\nu\nu'}\,.
\label{eq:mod3}
\end{equation}
The correlation between different frequency channels is then encoded in 
the matrix ${\bf R}$, which is assumed to be independent of the angular 
scales $\ell$. If the signal is uncorrelated, ${\bf R}$ reduces to the 
identity matrix (e.g., uncorrelated detector noises) while, for perfectly 
correlated signals, it is a rank--1 matrix containing only unit entries 
(e.g., CMB fluctuations). 
We expect that Galactic foregrounds typically fall into an intermediate 
case. Since we presently lack detailed measurements of the foreground 
correlation matrice ${\bf R}$, we decided to follow the simple model 
developed by \citet{teg98,teg00} and write the correlation matrix in 
terms of a parameter $\zeta$, the {\it frequency coherence}:
\begin{equation}
{\bf R}_{\nu\nu'}\simeq\exp\Bigg\{-\frac{1}{2}
\Bigg[\frac{\ln(\nu/\nu')}{\zeta}\Bigg]^2\Bigg\}\,.
\end{equation}
The frequency coherence determines the extent to which two frequencies 
can be separated before their correlation starts to break down. The two
limits $\zeta\to0$ and $\zeta\to\infty$ correspond to the two extreme
cases discussed above. For foregrounds with a spectrum such as the
dust emission, \citet{teg98} showed that the frequency coherence is of
the order of the inverse spectral index dispersion,
$\zeta\approx1/\sqrt{2}\Delta\beta$, where $\Delta\beta$ is the rms
dispersion of the dust emissivity index. \citet{planck_xxii_2015}
measured the mean value of $\beta$ at intermediate latitudes for
frequencies $\le353$\,GHz and found $\langle\beta\rangle=1.51$
with 1--$\sigma$ dispersion of 0.07. We shall hereafter use this value 
for $\Delta\beta$, which leads to a frequency coherence of $\zeta=10.1$. 
In the frequency range we are interested in, this guarantees a very high 
level of correlation, always larger than 99\%. We will investigate in
\S\ref{sec:discussion} the effects of having larger values of
$\Delta\beta$.

\subsection{Dust foreground removal}
\label{ssec:removal}
 
Contamination by Galactic dust emission is the strongest limitation
for measurements of CIB fluctuations, and accurate methods to separate 
it from the CIB signal are required. 
In this section however, we are not interested in the ability of a 
particular foreground subtraction method to separate dust from CIB. 
Instead, we shall hereafter assume that the Galactic dust removal can 
be done correctly down to a given level (e.g., 10\%, 1\%, etc.), so 
that we can investigate in which way the presence of foreground 
residuals propagate into the uncertainty on the $\fnl$ parameter.

The problem is equivalent to forecasting the detectability of the CMB
B--mode polarization and the tensor--to--scalar ratio $r$ parameter in
CMB experiments \citep[see, e.g.,][]{err15}. Analogously to the CMB,
CIB fluctuations can be treated as being statistically isotropic on 
the sky, yet with the important difference that the CIB frequency 
spectrum is not perfectly known and depends on the CIB model parameters.

Different authors \citep[e.g.,][]{tuc05,sti10,err15} have shown that,
after the subtraction, power spectra of foreground contaminants
leftover in CMB maps can be fairly well described by the original power
spectra scaled down by a factor that depends on details of the
component separation and properties of signals. Moreover, the
noise variance in the reconstructed CMB maps will be degraded according 
to the frequency spectrum of foregrounds and the noise in the channels
involved in the foreground subtraction. Assuming that the frequency 
dependence of the foregrounds is perfectly known, the noise variance in 
the reconstructed CMB maps -- obtained by a standard minimum--variance
solution -- is \citep[see, e.g.,][]{teg00,sto09,err15}:
\begin{equation}
\Sigma^2_{\CMB}=\bigg[({\bf A}^T{\bf N}^{-1}{\bf A})^{-1}
\bigg]_{\CMB\,\CMB}\,,
\label{eq:remov1}
\end{equation}
where ${\bf A}$ is the ``mixing'' matrix that describes the frequency
dependence of the sky signal components (i.e., CMB and foregrounds),
and has a dimension of $[N_{\nu}\times N_s]$ ($N_{\nu}$ is the number
of frequencies and $N_s$ the number of sky components). The square 
$[N_{\nu}\times N_{\nu}]$ matrix ${\bf N}$ is the noise covariance
matrix. Eq.\,(\ref{eq:remov1}) is a good approximation even when the 
spectral behavior of foregrounds is parametrized by a set of spectral 
parameters which need to be determined together with the sky signal 
\citep[see, however, the discussion in][]{sto09}.

We extend the previous formalism to the CIB, which is now the
component to be recovered. For simplicity, we assume that the sky
signal is composed only of Galactic dust emission and CIB
fluctuations. The elements ${\bf A}_{i1}$ and ${\bf A}_{i2}$ of the
[$N_{\nu}\times2$] mixing matrix at the frequency $i$ are then given
by Eqs.\ (\ref{eq:cibsed}) and (\ref{eq:mod1}), where the reference
frequency $\nu_0$ now is the frequency at which we want to recover the
CIB map. If the noise in the different channels is uncorrelated, the
noise variance in CIB reconstructed maps is
\begin{equation}
\Sigma^2_{\CIB}(\nu_0)=
\frac{\sum_i\Theta_d^2(i)/\sigma^2_i}
{{\rm det}({\bf A}^T{\bf N}^{-1}{\bf A})}\,,
\label{eq:remov2}
\end{equation}
with
\begin{align}
{\rm det}({\bf A}^T{\bf N}^{-1}{\bf A}) &=
\sum_i\frac{\Theta_{\CIB}^2(i)}{\sigma^2_i}\sum_i\frac{\Theta_d^2(i)}
{\sigma^2_i}\\ 
&\qquad -\Bigg[\sum_i\frac{\Theta_{\CIB}(i)\Theta_d(i)}{\sigma^2_i}
\Bigg]^2 \nonumber \,,
\nonumber
\end{align}
and $\sigma_i$ is the noise level at the $i$th frequency. We see that,
in the regime $\Theta_{\CIB}\sim\Theta_d$ expected at frequencies
$\lsim300$\,GHz (see Figure\,\ref{fig:sed}), the noise in the
reconstructed map diverges. Frequencies much larger than 300\,GHz are
therefore mandatory to separate CIB fluctuations and dust emission.

Following, e.g., \citet{err15}, we assume that the mixing matrix
${\bf A}$ can be directly estimated with a set of frequency maps
produced by one (or more) experiment. The reconstructed CIB map (at
our reference frequency $\nu_0$) will be then obtained by linearly
combining the set of frequency maps on the basis of the mixing
matrix. This can be seen as a general approach, independently of the
specific component separation method employed. However, unlike for the
CMB, we want to recover independent CIB maps at several frequencies. I
f observations are available at $n$ frequencies ($\nu\ga200$\,GHz), we 
assume that ``clean'' CIB maps can be reconstructed only in 
$n_{\CIB}\le n/2$ frequencies, while the other $n_d=n-n_{\CIB}$ channels 
are dedicated to the estimation of the mixing matrix and the Galactic 
dust template used in the subtraction. 
The noise variance in a recovered CIB map is then given by
Eq.\,(\ref{eq:remov2}), where the mixing matrix is evaluated using the
$n_d$ ``dust'' channels plus the CIB channel at $\nu_0$.

The auto-- and cross-- power spectra in clean CIB maps read
\begin{equation}
C_{\ell}^{\nu_i\nu_j}=C_{\ell}^{(\CIB)}(\nu_i,\nu_j)+\varepsilon 
C_{\ell}^d(\nu_i,\nu_j)+\Sigma_{\CIB}^2(\nu_i)\delta_{ij}\,,
\label{eq:remov4}
\end{equation}
where $\Sigma_{\CIB}$ is the noise variance obtained from
Eq.\,(\ref{eq:remov2}), and $\varepsilon$ is the fraction of the total
Galactic dust power spectrum leftover in CIB maps. We use as
reference value $\varepsilon=10^{-2}$, which implies a subtraction of
the dust emission at the level of 99\%. This level should be the
minimum goal for future space missions characterized by high
sensitivity and large frequency coverage. 

In the following forecasts, we will include two extra parameters in the
Fisher matrix (Eq.\,\ref{eq:fisher}): the amplitude $A_d$ of the dust
power spectrum and the spectral power--law index $\alpha_d$, shown
Eq.\ (\ref{eq:mod2}). $\sigma(\fnl)$ will be also marginalized over them.

\begin{figure}
\centering
\includegraphics[width=80mm]{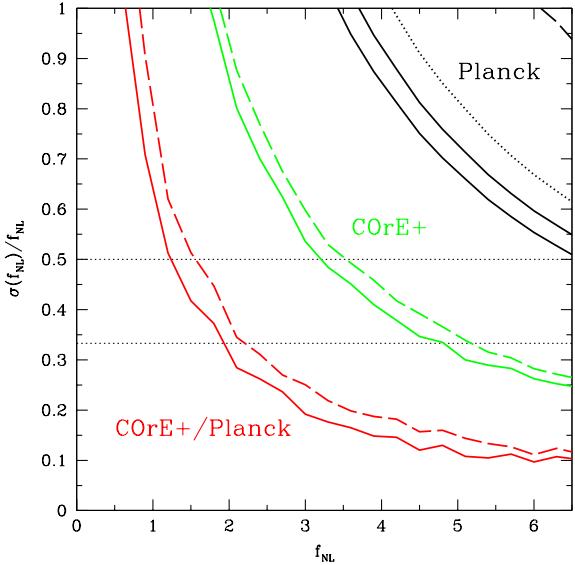}
\caption{The relative uncertainty for positive $\fnl$ obtained by
  Fisher forecasts for \planck\ (black lines), COrE+ (green lines) and
  a combination of COrE+ and \planck\ (red lines). The two black solid
  lines are for \planck\ with $\fsky = 0.2$ (upper) and 0.4 (lower
  curve; the case with $\fsky=0.6$ is very close to the case with 0.4); the
  dotted black line is for $\fsky=0.1$ and the dashed line (in the
  upper--right corner) for $\fsky=0.8$. For COrE+, the solid (dashed)
  green line corresponds to the configuration with 4 (3) CIB frequency
  maps. The red dashed line is for COrE+ adding only the 857--GHz
  channels, and the solid line adding all the {\it Planck} frequencies
  $\ge353$\,GHz (see the text and Table\,\ref{tab:3})}
\label{fig:resu1}
\end{figure}

\begin{figure*}
\centering
\includegraphics[width=55mm]{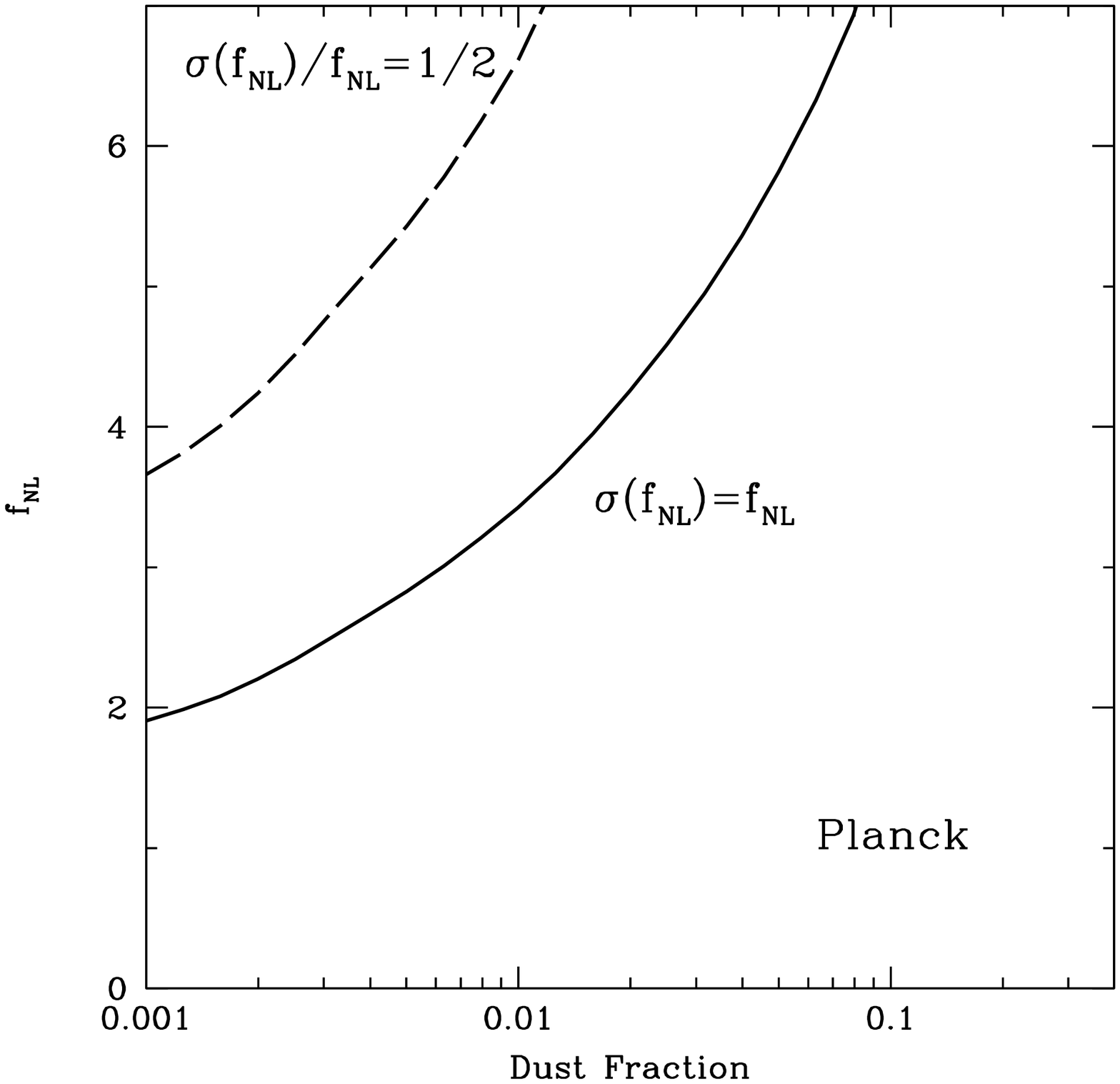}
\includegraphics[width=55mm]{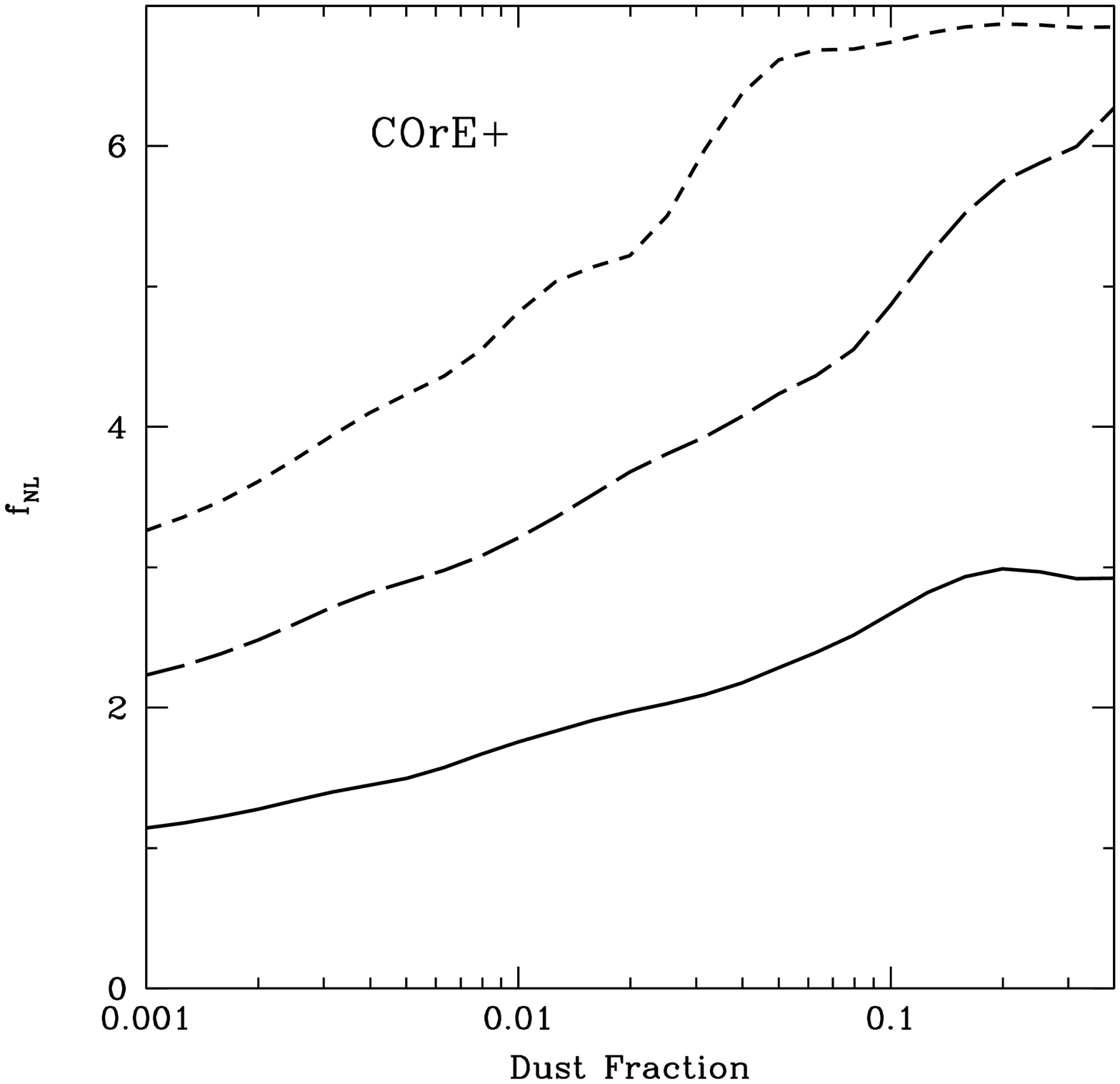}
\includegraphics[width=55mm]{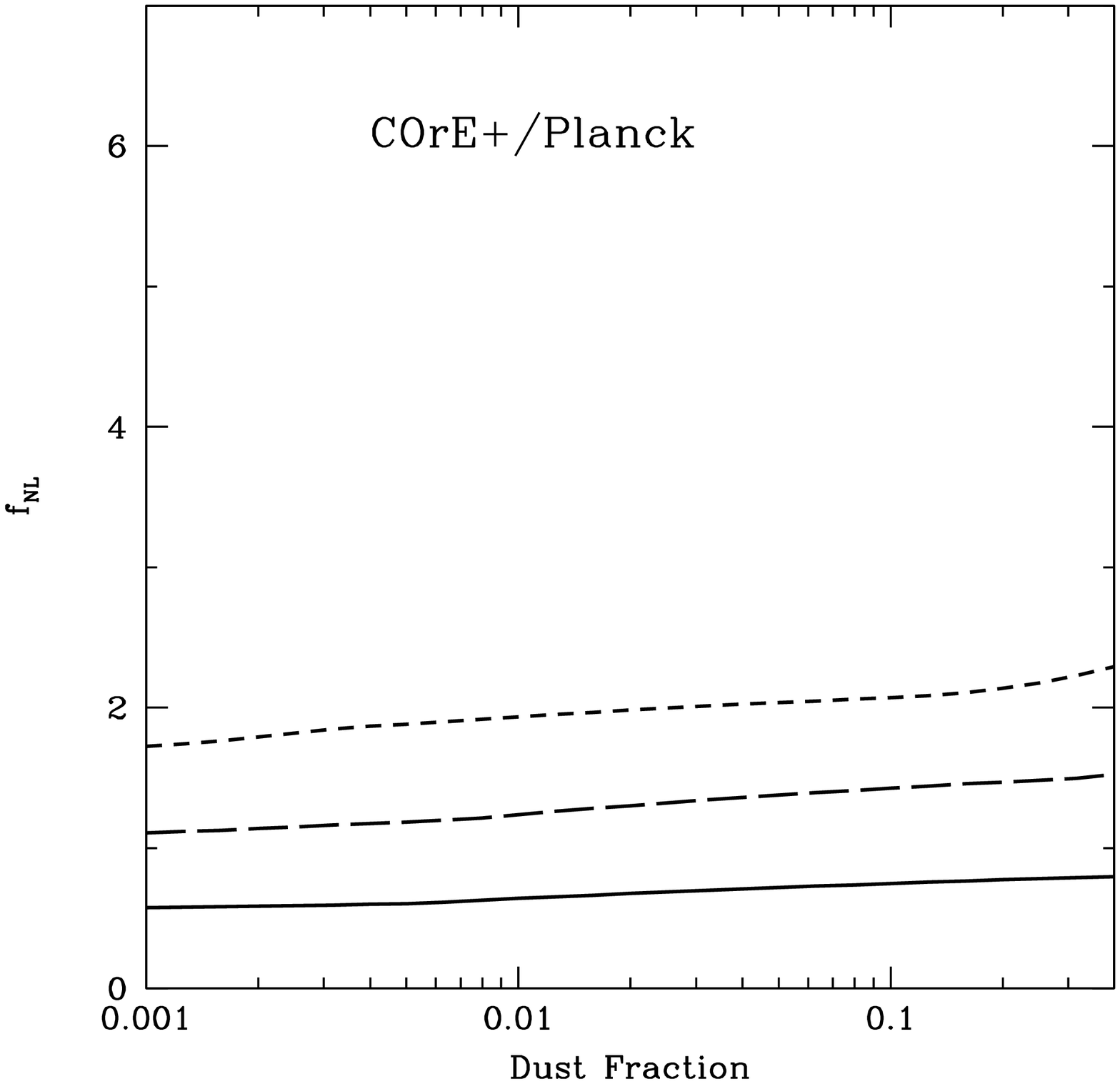}
\caption{The value of $\fnl$ for which its relative uncertainty is
  equal to 1 (solid lines, corresponding to a $1\sigma$ detection), 
  1/2 (long dased lines) and 1/3 (short
  dashed lines) is plotted as a function of the fraction of the dust
  residual. This is computed for \planck\ (left panel), COrE+
  (middle panel) and the combination COrE+/\planck\ (right panel),
  in their best configuration for CIB measurements.}
\label{fig:resu2}
\end{figure*}

\begin{figure}
\centering
\includegraphics[width=80mm]{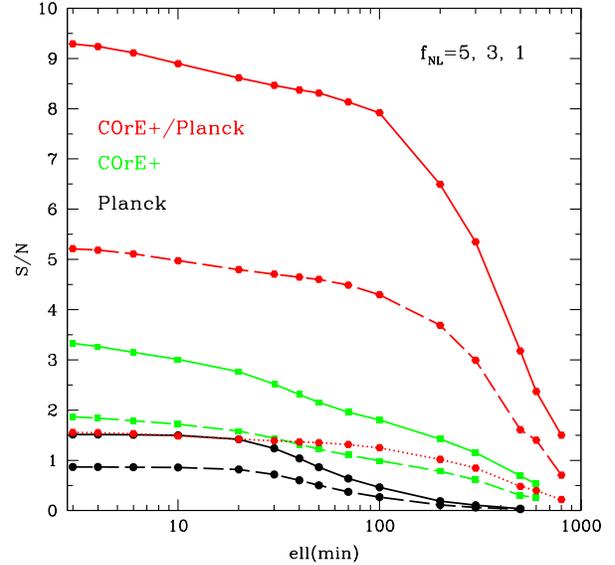}
\caption{Signal--to--noise ratio for $\fnl=5,$ 3 and 1 (solid, dashed
  and dotted lines, respectively) as expected for \planck\ (red
  lines), COrE+ (green lines) and the combination COrE+/\planck\
  (red lines). This is computed for $\fsky=0.4$ and 1\% dust residual.}
\label{fig:resu3}
\end{figure}

\subsection{Fisher forecasts for present and future experiments}
\label{ssec:results}

In this section we aim to provide realistic forecasts for the
detection of the $\fnl$ parameter including 1) the Galactic dust
contamination, as discussed above, and 2) the instrumental properties
of CMB space missions. We focus on \planck, and on possible
future experiments like COrE+, LiteBIRD \citep{litebird} and PIXIE
\citep{pixie}. Table\,\ref{tab:3} gives the instrumental
specifications we consider for these experiments. We take into account
frequencies $\nu>200$\,GHz solely.

{\bf Planck.} Observations from the \planck\ mission cover a large
range of frequencies, up to 857\,GHz. The 545-- and 857--GHz channels
are crucial for the CIB/dust separation. In our analysis we suppose 
that the channels at 353 and 857\,GHz are dedicated to provide the 
dust template, while CIB maps are recovered at 217 and 545\,GHz. In
Table\,\ref{tab:3} we report the degraded sensitivity in these
channels after the subtraction of the dust contamination (see
Eq.\,\ref{eq:remov2}). In this particular case, the 217--GHz channel 
gains, in term of (squared) sensitivity, a factor $\sim1.5$, while the 
545--GHz channel loses sensitivity by almost a factor 6.

Assuming that the dust emission has been cleaned at a level of 1\%, we
find that the uncertainty on $\fnl$ provided by \planck\ is
$\sigma(\fnl)\simeq3.5$, almost independently of the value of
$\fnl$. In Figure\,\ref{fig:resu1} we show the relative uncertainty on
$\fnl$. According to our estimates, \planck\ should be able to detect
$|\fnl|$ at about 2--$\sigma$ level if it is larger than 6, and give
an upper limit of $\sim3.5$ at 1--$\sigma$. This would significantly
improve the current constraints on the local $\fnl$ provided by
\planck\ using the CMB bispectrum \citep{planck_xvii_2015},
i.e. $\fnl=0.8\pm5.0$.

We have also considered CIB maps covering different fractions of the
sky. The optimal configurations have a sky fraction $\fsky\simeq0.4$, 
although similar results are obtained for $0.2\lsim \fsky\lsim0.6$. 
For smaller areas, the sensitivity to $\fnl$ gets worse, while very 
large fractions of the sky are not useful for the $\fnl$ detection due 
to the strong dust contamination. Finally, we show in Figure\,\ref{fig:resu2}
the dependence of our results on the amplitude of the dust residual. 
The possible detection of $\fnl$ strongly depends on the dust contamination. 
In order to improve the current \planck\ constraint on $\fnl$, dust 
contamination has to be removed at least at a 3\% level. 
On the other hand, a 1\,per\,mil cleaning would give an uncertainty of 
$\sigma(\fnl)\simeq2$.

{\bf COrE+.} This project represents an approximately optimal space
mission for the search of the CMB B--mode polarization, characterized
by a high sensitivity of a few $\mu$K--arcmin, a few arcmin resolution 
and a large frequency coverage. The COrE+ project is planned to include 
8 frequencies at $\nu>200$\,GHz, with the highest channel at
600\,GHz. We consider two possible configurations for the measurement
of the CIB: a more conservative choice involving 3 frequencies for CIB
and 5 for dust, and a optimistic one with half of the frequencies 
reserved for each component (see Table\,\ref{tab:3}). 
In both cases, after the component separation, the noise in the 
reconstructed CIB maps increases by a factor 5--10. 
This significant degradation in sensitivity may be related to the 
frequency coverage, which does not include the wavelengths at which the 
SEDs of the two components start to differ. 
As expected however, the sensitivity on $\fnl$ significantly increases 
with respect to \planck, being $\sigma(\fnl)\ga1.6$. 
Consequently, $|\fnl|=5$ would be measured at $\sim3$--$\sigma$, and 
$|\fnl|$ of 2--3 would be also detectable with a significance of about 
1--\,2\,$\sigma$ (see Figure\,\ref{fig:resu1}). 
The difference between the two configurations is modest, although an 
extra CIB frequency has been added. Here again, the best results are 
for sky fractions around 0.4--0.6. Looking at Figure\,\ref{fig:resu2}, 
the COrE+ results appear to be less dependent on the dust residual, 
presumably owing to the higher instrumental sensitivity.

{\bf COrE+/Planck.} The sensitivity of COrE+ to the $\fnl$ parameter
strongly improves when the highest \planck\ channels are also
taken into account and combined with COrE+ data. Extending the
frequency coverage to 857\,GHz is key to better separate CIB and
dust emission. It is actually enough to include the \planck\
857--GHz channel in the mixing matrix to reduce the noise degradation
in reconstructed CIB maps, at a level close to the COrE+ sensitivity
(see Table\,\ref{tab:3}). This is due to the different scaling of
  the dust and the CIB at high frequencies, see Fig.\,\ref{fig:sed}.
In Figure\,\ref{fig:resu1} we show the results when {\it Planck} data
are included to estimate the mixing matrix and the noise
degradation. We see that a 3-- (2--)$\sigma$ detection of $|\fnl|$ is
now possible down to values of 2 (1.5). 
This means that, with this configuration,
the residual dust emission can be accurately separated from the CIB
fluctuations. This is confirmed by the fact that reducing the level of
the dust residual there is a very small improvement in the detection
of $\fnl$ (see Figure\,\ref{fig:resu2}). In addition, if we increase
the fraction of the sky up to $\fsky=0.8$, $\sigma(\fnl)$ always
decreases.

Figure\,\ref{fig:resu3} shows the signal--to--noise ratio (defined as
${\rm SNR}=\fnl/\sigma(\fnl)$) for $\fnl$ as a function of the minimum
observable multipole $\lmin$ for the three cases discussed above. It
is interesting to note that for \planck\ the SNR is constant up to
$\lmin=20$, indicating that at the largest angular scales the signal
is completely dominated by the dust contamination. On the contrary,
COrE+ succeeds to better separate the two components and to access
information also at $\ell\lsim20$. The multipole range that
contributes the most to the SNR changes according to the experiment
and the configuration: in \planck\ this range is about $\ell=20$--70;
in COrE+ it extends almost up to $\ell=300$--400; in COrE+/\planck\
large angular scales seem to contribute less than in the previous
examples, and the SNR clearly drops only at $\ell>100$ and the main
contribution comes from scales between $\ell\sim100$ and $\lsim600$.

{\bf LiteBIRD.} This satellite could provide interesting constraints
on $\fnl$ in its extended
version\footnote{http://indico.cern.ch/event/506272/contributions/2138028/},
with several spectral bands spanning 40--400\,GHz and 4 bands between
200 and 400\,GHz. Clearly, without external information, LiteBIRD
alone is not suitable for detecting PNG through the CIB
($\sigma(\fnl)\ga6$, see Table\,\ref{tab:3}). However, jointly with
{\it Planck} data, we find $\sigma(\fnl)\simeq1.3$ (in a
configuration where three out of four LiteBIRD channels are dedicated 
to the CIB), only a factor 2 larger than with COrE+.

{\bf PIXIE.} The planned PIXIE mission presents almost the optimal
requirements for detecting $\fnl$ through CIB anisotropies, thanks to
its 400 channels between 30\,GHz and 6\,THz and a rms sensitivity of
the order of 70\,nK in $1^{\circ}$ pixels. In these conditions, dust
emission should be extremely well constrained, even at the peak of the
thermal emission. We expect therefore that PIXIE will be able to
investigate very tiny PNG signals, with $|\fnl|\ll1$. As an example,
we have considered 8 frequency bands between 200 and 600\,GHz with a
squared sensitivity of 1\,Jy\,sr$^{-1}$ and a resolution of
$1.6^{\circ}$: we obtain $\sigma(\fnl)\simeq0.08$ assuming 1\%
of dust residual and $\sim0.07$ without dust contamination. Extending
the frequency range to 900\,GHz (but with a similar number of
frequency bands) seems not to produce any relevant improvement in the
$\fnl$ uncertainty. However, these results should be taken as
indicative. On one hand, we expect that increasing the number of
frequency bands dedicated to CIB, $\sigma(\fnl)$ could still
decrease. On the other hand, especially for the PIXIE case that can
investigate very low values of $\fnl$, a more accurate analysis should
be required, taking into account effects of other contaminants (CMB,
extragalactic sources, etc.)  and model uncertainties (e.g., in the
CIB and bias prescriptions). We will discuss these issues in the next
Section.

\begin{table*}
\begin{center}
  \caption{Instrumental specifications of the space missions
    considered in the paper, the corresponding noise variance in the
    reconstructed CIB maps for different configurations ($\Sigma^2$)
    and the uncertainty on $\fnl$ (assuming $\fnl=0$). \label{tab:3}}
\begin{tabular}{ccccccccccccccccccc}
\hline
& \multicolumn{16}{c}{\it Planck} & & $\sigma(\fnl)$ \\
\hline 
$\nu$ [GHz] & & \multicolumn{2}{c}{217} & \multicolumn{2}{c}{} & 
\multicolumn{2}{c}{} & \multicolumn{3}{c}{353} &
\multicolumn{3}{c}{} & 545 &  & 857 & & \\
fwhm [arcmin] & & \multicolumn{2}{c}{5.02} & \multicolumn{2}{c}{} & 
\multicolumn{2}{c}{} & \multicolumn{3}{c}{4.94} &
\multicolumn{3}{c}{} & 4.83 &  & 4.64 & & \\
w$^{-1}$ [Jy$^2$\,sr${-1}$] & & \multicolumn{2}{c}{43.32} & 
\multicolumn{2}{c}{} & \multicolumn{2}{c}{} & \multicolumn{3}{c}{164.7} &
\multicolumn{3}{c}{} & 185.3 &  & 157.9 & & \\
\hline 
$\Sigma^2$ [Jy$^2$\,sr${-1}$] & & \multicolumn{2}{c}{29.0} & 
\multicolumn{2}{c}{} & \multicolumn{2}{c}{} & \multicolumn{3}{c}{--} &
\multicolumn{3}{c}{} & 1100 &  & -- & & {\bf 3.6} \\
\hline
\hline
& \multicolumn{16}{c}{COrE+}  & & \\
\hline
$\nu$ [GHz] & & \multicolumn{2}{c}{220} & \multicolumn{2}{c}{255} &
\multicolumn{2}{c}{295} & \multicolumn{3}{c}{340} &
\multicolumn{2}{c}{390} & 450 & 520 & 600 &  & & \\
fwhm [arcmin] & & \multicolumn{2}{c}{3.82} & \multicolumn{2}{c}{3.29} &
\multicolumn{2}{c}{2.85} & \multicolumn{3}{c}{2.47} &
\multicolumn{2}{c}{2.15} & 1.87 & 1.62 & 1.40 & & & \\
w$^{-1}$ [Jy$^2$\,sr${-1}$] & & \multicolumn{2}{c}{0.654} & 
\multicolumn{2}{c}{1.43} &
\multicolumn{2}{c}{5.20} & \multicolumn{3}{c}{8.31} &
\multicolumn{2}{c}{13.50} &  22.98 & 39.88 & 69.26 & & & \\
\hline
$\Sigma^2$ [Jy$^2$\,sr${-1}$] & & \multicolumn{2}{c}{--} & 
\multicolumn{2}{c}{8.03} &
\multicolumn{2}{c}{--} & \multicolumn{3}{c}{47.3} &
\multicolumn{2}{c}{--} & -- & 377. & -- &  & & {\bf 1.8} \\
& & \multicolumn{2}{c}{--} & \multicolumn{2}{c}{8.04} &
\multicolumn{2}{c}{--} & \multicolumn{3}{c}{47.3} &
\multicolumn{2}{c}{--} & 212. & 382. & -- &  & & {\bf 1.6} \\
with {\it Planck} 857\,GHz & & \multicolumn{2}{c}{--} & 
\multicolumn{2}{c}{1.83} &
\multicolumn{2}{c}{--} & \multicolumn{3}{c}{10.0} &
\multicolumn{2}{c}{--} & -- & 82.6 & 147. &  & & {\bf 0.7} \\
with {\it Planck} & & \multicolumn{2}{c}{--} & 
\multicolumn{2}{c}{1.97} &
\multicolumn{2}{c}{--} & \multicolumn{3}{c}{10.8} &
\multicolumn{2}{c}{--} & 45.6 & 90.2 & 163.6 &  & & {\bf 0.6} \\
\hline
\hline
& \multicolumn{16}{c}{LiteBIRD} & & \\
\hline
$\nu$ [GHz] & & \multicolumn{2}{c}{235} & \multicolumn{2}{c}{} & 
\multicolumn{2}{c}{280} & \multicolumn{3}{c}{337} &
\multicolumn{2}{c}{402} & & &  & & & \\
fwhm [arcmin] & & \multicolumn{2}{c}{30} & \multicolumn{2}{c}{} & 
\multicolumn{2}{c}{30} & \multicolumn{3}{c}{30} &
\multicolumn{2}{c}{30} & & &  & & & \\
w$^{-1}$ [Jy$^2$\,sr${-1}$] & & \multicolumn{2}{c}{0.36} & 
\multicolumn{2}{c}{} & 
\multicolumn{2}{c}{1.45} & \multicolumn{3}{c}{1.1} &
\multicolumn{2}{c}{0.7} & & & & & & \\
\hline
$\Sigma^2$ [Jy$^2$\,sr${-1}$] & & \multicolumn{2}{c}{11.3} & 
\multicolumn{2}{c}{} & 
\multicolumn{2}{c}{} & \multicolumn{3}{c}{149.5} &
\multicolumn{2}{c}{} & & &  & & & {\bf 6.4} \\
with {\it Planck}  & & \multicolumn{2}{c}{1.0} & 
\multicolumn{2}{c}{} & 
\multicolumn{2}{c}{} & \multicolumn{3}{c}{6.7} &
\multicolumn{2}{c}{14.6} & & &  & & & {\bf 1.3} \\
\hline
\hline
& \multicolumn{16}{c}{PIXIE} & & \\
\hline
$\nu$ [GHz] & & \multicolumn{2}{c}{218} & \multicolumn{2}{c}{248} &
\multicolumn{2}{c}{293} & \multicolumn{3}{c}{353} &
\multicolumn{2}{c}{398} & 443 & 548 & 593 & & & \\
fwhm [arcmin] & & \multicolumn{2}{c}{96} & \multicolumn{2}{c}{96} &
\multicolumn{2}{c}{96} & \multicolumn{3}{c}{96} &
\multicolumn{2}{c}{96} & 96 & 96 & 96 & & & \\
\hline
$\Sigma^2$ [Jy$^2$\,sr${-1}$] & & \multicolumn{2}{c}{1.0} & 
\multicolumn{2}{c}{1.0} & \multicolumn{2}{c}{1.0} & \multicolumn{3}{c}{1.0} &
\multicolumn{2}{c}{1.0} & 1.0 & 1.0 & 1.0 &  & & {\bf 0.08} \\
no dust & & \multicolumn{2}{c}{1.0} & 
\multicolumn{2}{c}{1.0} & \multicolumn{2}{c}{1.0} & \multicolumn{3}{c}{1.0} &
\multicolumn{2}{c}{1.0} & 1.0 & 1.0 & 1.0 &  & & {\bf 0.07} \\
\hline
\hline
\end{tabular}
\end{center}
\end{table*}

\section{Discussion}
\label{sec:discussion}




{\bf Model of CIB power spectra.} Our Fisher forecasts rely on the
fact that the CIB model we adopt provides a good description of actual
CIB spectra, at least on the angular scales of interest. This is a
reasonable hypothesis, considering that the model is able to fit
observations for auto-- and cross--frequency power spectra, number
counts, and absolute CIB levels simultaneously
\citep{vie13,planck_xxx_2013}. However, although it successfully
describes {\it Planck} data, it remains nonetheless a fairly simplified 
description of the actual CIB. Future high--quality CIB measurements 
(necessary to detect the $\fnl$ parameter) may require an additional 
level of sophistication. 

\begin{figure}
\centering
\includegraphics[width=80mm]{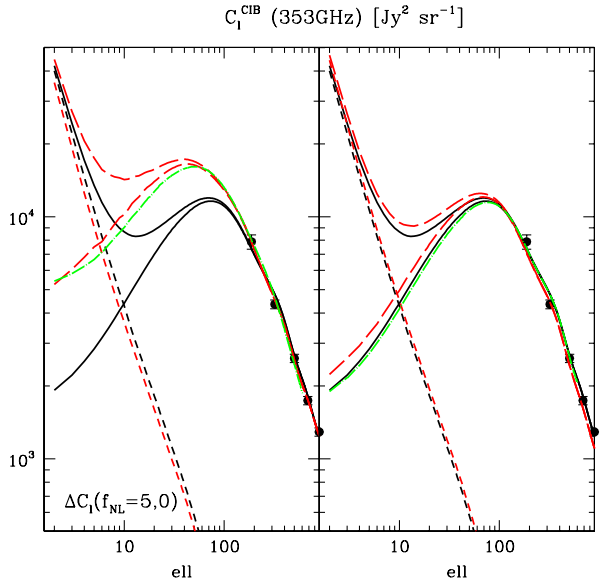}
\caption{{\it Left panel} CIB power spectra with $\fnl=0$ and 5,
  according to different model approaches: the extended halo model
  used in the paper (solid black lines); the simplified halo model
  (red long dashed lines); the linear model (green dash--dotted
  line, only $\fnl=0$). Short dotted lines shows the difference between the Gaussian case
  and $\fnl=5$. Points represent {\it Planck} measurements
  \citep{planck_xxx_2015}. CIB spectra differ at $\ell<200$. {\it
    Right panel} The same as in the left panel, but here spectra are
  computed using the same galaxy emissivity evolution as given by the
  extended halo model. Differences in CIB spectra are now small or
  negligible over all the multipole range considered. In either case, the PNG signal seems
  to be independent on the model approach.}
\label{fig:discu1}
\end{figure}

The CIB angular power spectra are currently well determined only at 
small/intermediate angular scales, i.e. $\ell\ga200$. 
No information is thus far available at larger scales, where the PNG 
effects on CIB anisotropies are most pronounced. 
Some concern could arise if the shape of the low--$\ell$ part of CIB 
spectra depends depends on model assumptions. To test this, we have 
compared the large--scale power spectra obtained with different CIB 
models, all providing a good fit to the {\it Planck} data. 
Besides the extended halo model used in this paper, we have considered: 
1) a linear model where the 2--halo contribution is determined by an 
effective bias and by the star formation history, which in turn are 
constrained by {\it Planck} data \citep[see \S5.4 of][]{planck_xxx_2013}; 
2) a simplified halo model in which all galaxies have the same luminosity 
regardless of their host dark matter halo used, e.g.. in 
\citet{planck_xviii_2011, cur15}. 
Although they agree at $\ell\ga200$, the power spectra exhibit a different 
behavior at the largest scales (see left panel of Figure\,\ref{fig:discu1}). 
Our reference model peaks at higher angular scales and has significantly
less power at $\ell\lsim100$. The main factor that determines the
low--$\ell$ behavior in the CIB spectra is the redshift evolution of the
average galaxy emissivity, $\bar\jmath_{\nu}(z)$. In fact, when the same 
$\bar\jmath_{\nu} (z)$ is applied to the different CIB models, the 
low--$\ell$ power spectra become very similar, and even indistinguishable 
(after a proper renormalization) as shown in the right panel of 
Figure\,\ref{fig:discu1}. 
Overall, the PNG features in the CIB power spectra seem to weakly depend
on the model prescription\footnote{In Figure\,\ref{fig:discu1}, for the linear model, we
  have considered only the Gaussian case. In fact, for this model it
  is not straightforward to introduce a NG bias due to the parametric
  approach used for the effective bias of infrared galaxies
  \citep[see][]{planck_xxx_2013}.}.

In order to distinguish PNG signals, it is therefore key to correctly
model the evolution of the infrared galaxy emissivity. Accurate
measurements of CIB power spectra at degree scales should further
strengthen the model reliability, especially concerning this
point. Estimates of number counts, luminosity and correlation
functions of infrared galaxies at shorter wavelengths should also
provide complementary information. In addition, as shown in
\citet{planck_xxx_2015}, the evolution of the galaxy emissivity can be
related to the star formation density history. The linear model and
the extended halo model recover consistent star formation histories
below $z=2$, in agreement with recent measurements by {\it Spitzer}
and {\it Herschel}, while at higher redshifts, there are discrepancies
between the two models and between models and observational estimates
\citep[see discussion in][]{planck_xxx_2015}. In the future, we expect
that a better determination of the history of the star formation
density will improve significantly the constraints on the galaxy
emissivity and then will help to fix the CIB spectral behavior at low
$\ell$s.

Moreover, we have shown in \S\ref{ssec:png} that the relative
amplitude of the non--Gaussian CIB bias could depend on the HOD
parameter $M_\text{eff}$, i.e. on the range of halo masses that are
most efficient at hosting star formation. In the extended halo model
this parameter is assumed to be constant in time, in agreement with
results from \citet{beh13a}, at least out to $z=4$. However, they
found a much lower value for the characteristic halo mass
(i.e. $\log(M_\text{eff}/M_{\odot})=11.7$) compared to the {\it
  Planck} results. We have verified that, using this value for
$M_\text{eff}$, CIB spectra change in amplitude but not in shape, even
with $\fnl\ne0$. This is an important indication that the PNG signal
is fairly insensitive to the HOD parameters. Note, however, that we
have extrapolated the subhalo mass function of \cite{tin10b} much
beyond its actual range of validity, i.e. $0.8<z<1.6$. Furthermore,
while the amplitude non-Gaussian bias is generally given by derivative
of $\bar n_h$ relative to the normalisation $\sigma_8$
\citep{Slosar:2008hx}, we have assumed a universal halo mass function
$\bar n_h$, which implies $\partial\ln\bar
n_h/\partial\ln\sigma_8=\delta_c b_1$. The validity of this assumption
may, however, depend on the halo identification algorithm
\citep{desjacques/seljak:2010}. Although N-body simulations support
universality when the mass of SO halos is $M\gtrsim 10^{13}M_\odot$
\citep{des09}, the extent to which this assumption holds at smaller
masses is still a matter of debate.

{\bf Galactic dust emission.} Due to its relevance, we should investigate
the extent to which our results are sensitive to the Galactic dust model 
used in the analysis. Dust residual in CIB reconstructed maps is parametrized --
see Section\,\ref{ssec:mod} and Eq.\ (\ref{eq:mod3}) -- by the product
of spatial, frequency and correlation terms. In particular, by taking
the correlation matrix $R_{\nu\nu'}$ independent of the angular scale,
we are supposing that the shape of dust residual power spectra are the
same over all the frequencies. This is something expected and
supported by simulations of component separation performance
\citep{cur16}. 

Two free parameters are considered for dust residual power spectra and
then marginalized in the Fisher approach: the amplitude and the
power--law index of the power spectrum at the reference frequency. The
dust SED is instead assumed to be perfectly known. This is motivated
by the fact that this Galactic contaminant largely dominates the sky
emission at $\nu\ga300$\,GHz, especially on large angular scales, and
the frequency spectrum is expected to be accurately determined across
the sky. This hypothesis can be relaxed by adding as extra free
parameter the dust emissivity index $\beta_d$. We have verified
however that it has a negligible impact on the $\fnl$ uncertainty.

A more critical issue in the Fisher analysis is the frequency
coherence in the dust residual. We have related it to the spatial
dispersion of the dust emissivity index, according to the prescription
proposed by \citet{teg98}. \citet{planck_xxii_2015} found that the
variance across the sky of the dust spectral index is
$\Delta\beta\simeq0.07$ which corresponds to a frequency correlation
level larger than 99\% (between 200 and 900\,GHz). It is clear that
reducing the frequency coherence of the dust emission can
significantly degrade the sensitivity on the $\fnl$ parameter for
experiments covering a large frequency range. In Table\,\ref{tab:4} we
compute the frequency coherence values as increasing $\Delta\beta$. We
can see that dust emission is still highly correlated for
$\Delta\beta\simeq0.1$, even between well separated frequencies, while
some decorrelation is observed for $\Delta\beta\sim0.3$ and
$\nu/\nu'>2$. The uncertainty on $\fnl$ from {\it Planck} or COrE+
seems to be quite sensitive to the frequency coherence and increases
by $\sim20$\% if $\Delta\beta=0.1$. On the contrary, when {\it Planck}
and COrE+ are combined, the frequency decorrelation has little impact
on results.
\begin{table}
\begin{center}
  \caption{Frequency coherence of the dust emission and the
    uncertainty on the $\fnl$ parameters as a function of the spatial
    dispersion of the dust spectral index. \label{tab:4}}
\begin{tabular}{ccccc}
\hline
$\Delta\beta$ & 0.07 & 0.1 & 0.3 & 0.5 \\
\hline
$\nu/\nu'$ & \multicolumn{4}{c}{$R_{\nu\nu'}$} \\
\hline
1.5 & $>0.99$ & $>0.99$ & 0.985 & 0.96 \\
2 & $>0.99$ & $>0.99$ & 0.96 & 0.89 \\
3 & $>0.99$ & 0.99 & 0.90 & 0.74 \\
4 & 0.99 & 0.98 & 0.84 & 0.62 \\
\hline
\hline
& \multicolumn{4}{c}{$\sigma(\fnl)$} \\
\hline
{\it Planck} & 3.6 & 4.2 & 7.5 & 9.1 \\
{\it COrE+} & 1.6 & 1.9 & 2.9 & 3.3 \\
{\it COrE+/Planck} & 0.6 & 0.64 & 0.72 & 0.76 \\
\hline
\end{tabular}
\end{center}
\end{table}
%

{\bf Dust contamination and Fisher forecasts.} As an alternative
approach to estimate $\fnl$ constraints in the presence of dust
contamination, we compute the Fisher matrix without assuming any
foreground removal. We include in the analysis all the observational
frequencies for a given experiment at their nominal sensitivity. We
expect this approach to produce more conservative results. Dust
removal on maps should be in fact more powerful and effective than a
CIB--dust emission separation at the level of power spectra. It is
also an useful test to check if our working hypothesis (e.g., dust
removal at a 1\,per\,cent level in about half of the observational
frequencies) are too conservative or too optimistic. For {\it Planck}
we find $\sigma(\fnl)\simeq5$, about a 50\,per\,cent larger than
previous results with 1\% dust residual. In this case, a component
separation approach to remove the dust contamination is strictly
required. On the other hand, in the case of COrE+, if we consider all
the 8 frequencies between 200 and 600\,GHz at their nominal
sensitivity, the uncertainty on $\fnl$ significantly decreases with
respect to previous values, and we find $\sigma(\fnl)\simeq0.25$ (to
be compared with $\sim1.6$ and 0.6 for COrE+ and COrE+/{\it Planck},
respectively). This result is very close to the one obtained without
any dust contamination in \S\ref{sec:fish}. Similarly, for LiteBIRD we
obtain $\sigma(\fnl)\simeq0.33$. Such an improvement can be explained by
the very small noise level expected in future experiments (not
degraded by component separation) together with a larger number of
frequency channels. We therefore conclude that, in principle, the dust
contamination should be fully controlled by high--sensitive
observations at a sufficient number of frequencies. Previous results for
COrE+ and LiteBIRD mission might be then conservative, at least in
terms of the number of frequencies to be exploited for the CIB.

For the purpose of computing the Fisher matrix we fixed the
  fiducial cosmological model.
  This assumption contributes to a possible theoretical systematic
  uncertainty. However, the \planck\ satellite has measured the
  parameters of the cosmological standard model to high precision
  \citep{planck_xiii_2015}. For example, the tilt of the primordial
  spectrum might be partially degenerate with the CIB slope at low
  $\ell$.  But \planck\ has determined the value of the scalar
  spectral index with tiny error bars, $n_s = 0.968 \pm 0.006$.  In
  addition, the \planck\ data covers a wide range of scales, and there
  is no indication for a running of the scalar spectral index or any
  other significant deviation from a power law spectrum, except maybe
  a (not significant) power deficit on large scales
  \citep{planck_xx_2015}. The power deficit can hardly be confused with the
  expected excess power on the largest scales in the CIB from
  non-Gaussian bias if $\fnl>0$, and also the curves for $\fnl<0$ turn
  back up at the lowest $\ell$ so that we do not expect this to be a problem.

Finally, we did not fully take into account the non-Gaussianity of
the CIB covariance matrix, although the CIB power spectrum includes
the non-Gaussian bias. We have ignored the contribution induced by
a primordial four-point function (or trispectrum) which, in the 
PNG model considered here, is proportional to $\fnl^2$. This term
might not be negligible at low multipoles. The covariance will also
include contributions generated by the nonlinear gravitational 
evolution. They will involve, among others, the CIB bispectrum.
However, these terms scale at best like $k^2$ in the power of 
biased tracers. Therefore, we expect them to contribute much less 
than the 1-halo term at low multipoles.


\vspace{3mm}

Overall, $\sigma(\fnl)<1$ results should be taken with some
caution. For example, according to our estimates, PIXIE has the
potentiality to detect PNG signals from $\fnl<0.1$. However,
as discussed in this section, our Fisher forecasts are based on some
theoretical assumptions (e.g., the bias prescription, Gaussian
covariance matrix, etc.), and on a particular CIB modeling framework,
whose validity and impact on results has not been fully
investigated. Finally, possible systematics in future data and extra
astrophysical contaminants (as CMB, point sources, thermal
Sunyaev--Zeldovich effect) have not been considered in the present
analysis. All these aspects should be investigated in future works.

\section{Conclusion}
\label{sec:conclusion}

In this paper we investigate the ability of CIB observations at
frequencies of a few hundred GHz (i.e.\ where CMB experiments operate)
to detect local primordial non-Gaussianity, and measure the nonlinear 
parameter $\fnl$ by leveraging the scale dependent bias on large scales.

%

We model the CIB angular power spectrum under the assumption that the 
cosmic infrared background is produced by star--forming galaxies emitting 
at infrared wavelengths. We describe the clustering of these galaxies 
using a halo model, with a galaxy luminosity--halo mass relation as adopted 
in \citet{sha12,planck_xxx_2013}.
We take into account both general relativistic corrections and the 
scale--dependent non--Gaussian halo bias induced by the local primordial
non-Gaussianity.
We perform a Fisher forecast to ascertain the precision at which $\fnl$
can be measured by CMB space missions.


We find that the GR corrections are subdominant except at the largest 
angular scales. The leading contribution arises from the Doppler term,
although it only makes up less than 50\% of the CIB angular power
spectrum at 353 GHz for $\ell=2$ (where cosmic variance is largest),
higher frequencies. At $\ell = 10$, the GR corrections are reduced to the 
percent level, corresponding to a local bispectrum shape with amplitude
$\fnl\simeq0.1$.

%

Under the ideal condition of perfect dust subtraction over 40\% of the 
sky, future CMB space missions like COrE+, operating at frequencies 
between 200 and 600\,GHz, shall be able to detect $\fnl\sim 1$ with high 
significance ($>2\,\sigma$), when several frequencies are combined. 
In this case, information from the very large angular scales, i.e. the 
ones most affected by Galactic dust, is not strictly required.
%


Of course, a complete subtraction of dust is unlikely. Therefore, we have
also produced forecasts assuming a dust subtraction at the 1\% level in
the best scenario.
In this case, an experiment with a \planck --like sensitivity should be 
able to reach an uncertainty on $\fnl$ of $\sigma(\fnl) \sim 3.5$ for sky
fractions between 0.2 and 0.6. Larger sky fractions are not useful because 
of the strong dust contamination, which actually degrades the constraints. 
Future probes like COrE+, LiteBIRD and PIXIE, which have a higher sensitivity 
and more frequency channels, should be able to constrain $\fnl$ to much lower 
values, even in a situation where the lowest multipoles could not be used 
owing to foreground contamination, systematic effects, or a residual dust
fraction higher than expected after the cleaning.
We have found that the presence of high-frequency channels in the range of 
800 to 1000 GHz is especially important to reduce the impact of dust and 
reach $|\fnl|<1$. 
Future missions should thus either include detectors at those frequencies 
or,alternatively, combine their data with the \planck\ 857 GHz to achieve
this precision.


To conclude, an analysis of current CIB observations should in principle 
already yield bounds on the local $\fnl$ competitive with those obtained 
from the analysis of non-Gaussianity in the \planck\ CMB maps.
The CIB holds the promise of reaching $|\fnl|\lsim 1$ in a not so distant
future. Although not considered here, additional information could also be
harvested from a cross-correlation of the ``CMB lensing potential'' and
CIB anisotropies. This is expected to be less sensitive to the PNG but, at 
the same time, less affected by dust contamination (Anthony Challinor, private
communication). While direct observations of the large scale structure might
eventually prove even more sensitive, the CIB will remain an invaluable tool 
to cross-check detections or limits from LSS due to the different systematics 
present in these various probes.

\section*{Acknowledgments}
It is a pleasure to thank Camille Bonvin and Anthony Challinor  
for helpful discussions. MT is also thankful to Enrique
Mart\'{\i}nez--Gonz\'alez and Patricio Vielva for useful discussions
about Fisher matrix forecasts. 
VD and MK are grateful to the Galileo Galilei Institute and the organizers
of the workshop on ``Theoretical Cosmology in the Era of Large Surveys'',
during which part of this work was completed. 
The authors acknowledge support by the Swiss National Science Foundation.
Part of the analysis was performed on the {\small BAOBAB} cluster at the 
University of Geneva.

\appendix

\section{General Relativistic corrections to the CIB intensity}

We provide details about the calculation of the GR correction to the CIB brightness. 
We refer the reader to the intensity mapping studies of \cite{Hall:2012wd,Alonso:2015uua} for
complementary information.

\subsection{CIB brightness}
\label{app:brightness}

Let $I(\nu_o)$ be the observed CIB brightness (in units of Jy sr$^{-1}$), i.e. the amount of 
energy $\dd E_o$ that passes through a proper surface element $\dd A_o$ (transverse to the observed 
ray direction) per frequency interval $\dd\nu_o$, per proper time $\dd\tau_o$ and per solid angle 
$d\Omega_o$,
\begin{equation}
I(\nu_o) = \frac{\dd E_o}{\dd\tau_o \dd\nu_o \dd\Omega_o \dd A_o} \;.
\end{equation}
The corresponding number of photons received by the observer is $\dd N_o=\dd E_o/h\nu_o$. 
Conservation of photon number (which follows from Liouville's phase space conservation law)
implies $\dd N_o=\dd N_s$, where $\d dN_s$ is the infinitesimal number of photons emitted by the 
source. Namely,
\begin{align}
I(\nu_o) &= \frac{\nu_o}{\nu_s} \frac{\dd E_s}{\dd\tau_s \dd\nu_s\dd\Omega_s \dd A_s} \,
\frac{\dd\tau_s \dd\nu_s}{\dd\tau_o \dd\nu_o}\,\frac{\dd\Omega_s \dd A_s}{\dd\Omega_o \dd A_o} \\
&= \left(1+z\right)^{-3} \frac{\dd E_s}{\dd\tau_s \dd\nu_s\dd\Omega_s \dd A_s} \nonumber \;.
\end{align} 
To derive the second equality, we have used $\nu_o/\nu_s=\dd\tau_s/\dd\tau_o=(1+z)^{-1}$ and the
reciprocity relation $D_L=(1+z)D_A$, where $D_A=\sqrt{\dd A_s/\dd\Omega_o}$ and 
$D_L=\sqrt{\dd A_o/\dd\Omega_s}$ 
are the angular diameter distance and (redshift-corrected) luminosity distance, respectively.
Note that the luminosity and angular distances have canceled each other out of the last 
expression. 

Let $\lambda$ be the affine parameter of the light ray $x^\mu(\lambda)$, so that 
$k^\mu=\dd x^\mu/\dd\lambda$ is its wavevector. 
During an interval $d\lambda$, the wavefront seeps out a proper volume $\dd V_s=\dd l_s\dd A_s$, 
where $\dd l_s=k_\mu u^\mu_s\dd\lambda$ is proportional to the 4-velocity $u_s^\mu$ of the emitter. 
Therefore, the above relation becomes
\begin{align}
\label{eq:IoJs}
I(\nu_o) &= 
\left(1+z\right)^{-3}\,\frac{\dd E_s}{\dd\tau_s \dd\nu_s \dd\Omega_s \dd V_s}\, 
(k_\mu u^\mu_s) \dd\lambda \\
&= \left(1+z\right)^{-3} \jmath_{\nu_s}^\text{\tiny phys}\,(k_\mu u_s^\mu) \dd\lambda \nonumber \;,
\end{align}
where
\begin{equation}
\jmath_{\nu_s}^\text{\tiny phys} = \frac{\dd E_s}{\dd\tau_s \dd\nu_s \dd\Omega_s \dd V_s}
\end{equation}
is the source emissivity per {\it physical} volume (in unit of Jy sr$^{-1}$ m$^{-1}$). 
In the case of a continuous medium, Eq.\ (\ref{eq:IoJs}) must be integrated along the light ray. 
Hence, we can write
\begin{equation}
\label{eq:Inuo}
I(\nu_o,\nvh) = \int\!\!\dd\lambda\, (k_\mu u_s^\mu)\, 
\frac{\jmath_{(1+z)\nu_o}^\text{\tiny phys}\!(\lambda)}{(1+z)^3} \;,
\end{equation}
where $\nvh$ is the observed angular position.
In principle, we should also take into account the loss of intensity as the photons propagate through 
the intergalactic medium and modify the above relation accordingly:
\begin{equation}
I(\nu_o,\nvh) = \int\!\!\dd\lambda\, (k_\mu u_s^\mu)\, 
\frac{\jmath_{(1+z)\nu_o}^\text{\tiny phys}\!(\lambda)}{(1+z)^3}e^{-\tau(\nu_o,\nvh,\lambda)} \;,
\end{equation}
where $\tau(\nu_o,\nvh,\lambda)$ is the optical depth along the path of the ray. For simplicity however, 
we will ignore the beam absorption in what follows, as the focus is the evaluation of Eq.\ (\ref{eq:Inuo}) 
at linear order in metric perturbations.

\subsection{GR corrections at first order in perturbations}
\label{app:GR}

We consider a perturbed FRW metric in the conformal Newtonian gauge with coordinates $x^\mu=(\eta,x^i)$, 
\begin{equation}
\dd s^2 = a^2(\eta) \big(\eta_{\mu\nu}+h_{\mu\nu}\big)\dd x^\mu \dd x^\nu \;,
\end{equation}
where
\begin{equation}
h_{00}=-2\Psi\;,\qquad h_{0i}=0\;,\qquad h_{ij}=-2\Phi\delta_{ij} \;.
\end{equation}
The perturbed photon wavevector $k^\mu=\dd x^\mu/\dd\lambda$ corresponding to the light ray $x^\mu(\lambda)$ 
can be written as
\begin{equation}
k^\mu = \frac{1}{a^2}\big(-1+\delta\nu,\nvh+\delta\nvh\big) \;,
\end{equation}
where $\delta\nu$ is the fractional frequency perturbation and $\delta\nvh$ is the change in the propagation 
direction, which is such that $\nvh^2=1$ and $\nvh\cdot\delta\nvh=0$.
Since lightlike geodesics are conformally invariant, the metric $(\eta_{\mu\nu}+h_{\mu\nu})\dd x^\mu\dd x^\nu$
admits the same photon geodesics which we parametrize as $x^\mu(\tilde\lambda)$. The corresponding photon
wavevector $\tilde k^\mu=\dd x^\mu/\dd\tilde\lambda$ is related to $k^\mu$ through
\begin{equation}
\tilde k^\mu = a^2 k^\mu \;.
\end{equation}
Hence, the two affine parameters satisfy $\lambda=a^2\tilde\lambda$ up to an irrelevant constant. 


We calculate the change in the perturbed photon wavevector $\tilde k^\mu$ using the geodesic equation 
\begin{equation}
\frac{\dd}{\dd\tilde\lambda} \tilde k^\mu = -\Gamma^\mu_{\alpha\beta} \tilde k^\alpha \tilde k^\beta\;,
\end{equation}
where $\Gamma^\mu_{\nu\lambda}$ are the Christoffel symbols in the conformally transformed metric. 
At linear order in perturbations, only the first geodesic equation describing the frequency shift is relevant 
as lensing of the uniform CIB yields the same CIB. 
Nonetheless, lensing magnification induces a first order perturbation to the CIB brightness through the flux
limiting value. We will discuss this effect below.
The relevant Christoffel symbols are $\Gamma^0_{00}=\dot{\Psi}$, $\Gamma^0_{0i}=\Psi_{,i}$ 
and $\Gamma^0_{ij}=-\dot{\Phi}\delta_{ij}$, where $\dot{X}$ and $X_{,i}$ designate a derivative w.r.t. $\eta$ 
and $x^i$, respectively. Linearizing the first geodesic equation, we arrive at
\begin{equation}
\label{eq:firstgeo}
\frac{\dd}{\dd\tilde\lambda}\delta\nu = 2\frac{\dd\Psi}{\dd\tilde\lambda}+\dot{\Psi}+\dot{\Phi} 
\end{equation}
upon taking advantage of the fact that the derivative w.r.t. $\tilde\lambda$ is
\begin{equation}
\frac{\dd}{\dd\tilde\lambda} \equiv \tilde k^\mu\partial_\mu \approx -\partial_\eta+\nh^i\partial_i
\end{equation}
at first order in perturbations.
Eq.\ (\ref{eq:firstgeo}) easily integrates to
\begin{equation}
\delta\nu(\lambda) - \delta\nu_o = 2\big(\Psi(\lambda)-\Psi_o\big)+\int_0^\lambda\!\!\dd\lambda'
\big(\dot{\Psi}+\dot{\Phi}\big)\;.
\end{equation}
It is now expressed in terms of the affine parameter $\lambda$, which we assume to be $\lambda=0$ at the observer
position. Furthermore, $\Psi_o=\Psi(\lambda=0)$, $\delta\nu_o=\delta\nu(\lambda=0)$ etc.

Assuming that the source at affine parameter $\lambda$ (a galaxy contributing to the CIB) and the observer 
move with a 4-velocity $u^\mu=a^{-1}(1-\Psi,v^i)$ and $u_o^\mu=a_o^{-1}(1-\Psi_o,v_o^i)$, where $v_i$ is 
the proper peculiar velocity, the redshift of the emitter relative to the observer (which could be measured 
if e.g. emission lines were resolved) is
\begin{equation}
1+z(\lambda) = \frac{\big(a^{-2} u_\mu \tilde k^\mu\big)_\lambda}{\big(a^{-2}u_\mu \tilde k^\mu\big)_o} 
\equiv \frac{1+\delta z(\lambda)}{a\big(\eta(\lambda)\big)}\;,
\label{eq:1+z}
\end{equation}
where $1/a\big(\eta(\lambda)\big)$ is the source redshift in an unperturbed universe, and 
\begin{equation}
\label{eq:dz}
\delta z(\lambda) = -(\Psi-\Psi_o) -\int_0^\lambda\!\!\dd\lambda'\,\big(\dot{\Psi}+\dot{\Phi}\big)
+(\vv-\vv_o)\cdot\nvh
\end{equation}
is the redshift perturbation.
Note that the boundary terms at the observer position generate an unmeasurable monopole ($-\Psi_o$) and a 
dipole 
($\vv_o\cdot\nvh$), which we will ignore in what follows. 

Since the emissivity $\jmath_\nu^\text{\tiny phys}$ explicitly depends on the observed redshift through 
$\nu=(1+z)\nu_o$, it is convenient to re-parametrize the light ray as a function of $z$, so that
\begin{equation}
\label{eq:Inuo_1}
I(\nu_o;\nvh) = \int\!\!\dd z\, \left(k_\mu u^\mu\frac{\dd\lambda}{\dd z}\right)\, 
\frac{\jmath_{(1+z)\nu_o}^\text{\tiny phys}\!\!(z,\vx(z))}{(1+z)^3} \;.
\end{equation}
Therefore, fluctuations are now defined at the hypersurfaces of constant observed redshift $z$. 
We begin by evaluating the term $k_\mu u^\mu (\dd\lambda/\dd z)$. 
Using $k_\mu u^\mu=a^{-1}(\eta)(1+z)$ with $z(\lambda)$ given by Eq.\ (\ref{eq:1+z}), we obtain
\begin{equation}
k_\mu u^\mu \frac{\dd\lambda}{\dd z} = \frac{a^2(\eta(z))}{{\cal H}(\eta(z))}
\left(1+\delta\nu - \frac{a^2(\eta(z))}{{\cal H}(\eta(z))}\frac{\dd\delta z}{\dd\lambda}\right)\;.
\end{equation}
We must now take into account the fact that the coordinate time fluctuates at fixed observed 
redshift $z$. Writing $\eta(z) = \bar\eta(z) + \delta\eta$, where $\bar\eta$ is the conformal
time corresponding to the observed redshift $z$ in the unperturbed background, the perturbation 
to $a^2/{\cal H}$ reads
\begin{equation}
\frac{a^2(\eta)}{{\cal H}(\eta)}=\frac{a^2(z)}{{\cal H}(z)}
\left[1+\left(2{\cal H}-\frac{\dot{\cal H}}{{\cal H}}\right)\delta\eta\right]\;,
\end{equation}
where $a(z) \equiv a(\bar\eta(z))$, ${\cal H}(z)\equiv {\cal H}(\bar\eta(z))$ and the overdot 
designates a derivative w.r.t. $\bar\eta$.
The time perturbation $\delta\eta$ can be read off from Eqs.~(\ref{eq:1+z}) and (\ref{eq:dz}) as
\begin{equation}
\label{eq:deta}
{\cal H}\delta\eta = -\Psi - \int_0^{\lambda(z)}\!\!\dd\lambda'\big(\dot{\Psi}+\dot{\Phi}\big)
+\vv\cdot\nvh \;.
\end{equation}
Here, $\lambda(z)$ is the value of the affine parameter at the observed redshift $z$.
Therefore, with the help of the relation ${\cal H}\delta\eta+\delta\nu = \Psi+\vv\cdot\nvh$, we eventually 
arrive at
\begin{equation}
k_\mu u^\mu \frac{\dd\lambda}{\dd z} = \frac{a^2(z)}{{\cal H}(z)} 
\left(1+\delta_\parallel\right)
\end{equation}
where
\begin{align}
\delta_\parallel &=
-\bigg(\frac{\dot{\cal H}}{\cal H}-{\cal H}\bigg)\delta\eta+\Psi+\vv\cdot\nvh \\
&\qquad +\frac{a^2(z)}{{\cal H}(z)}
\bigg[\frac{\dd\Psi}{\dd\lambda}+\big(\dot{\Psi}+\dot{\Phi}\big)
-\frac{\dd\vv}{\dd\lambda}\cdot\nvh\bigg] \nonumber
\end{align}
is the fractional change in $\dd l_s$, i.e. in the source volume element along the line of sight.
Note that the second term in the square brackets can be evaluated using
$(\dd/\dd\tilde\lambda)=a^2(\dd/\dd\lambda)$.

We now turn to the emissivity $\jmath^\text{\tiny phys}_\nu$. With the time slicing adopted here, 
perturbations to the emissivity are defined at constant observed redshift $z$:
\begin{equation}
a^3\jmath_\nu^\text{\tiny phys}(z,\vx) = 
a^3(z)\,\bar\jmath_\nu^\text{\tiny phys}(z)\Big(1+\delta_\jmath^z(z,\vx)\Big) \;,
\end{equation}
where it is understood that all quantities are evaluated at a frequency $\nu=(1+z)\nu_o$. In practice, 
it is useful to relate the (gauge-invariant) fluctuation $\delta_\jmath^z$ to the perturbation 
$\delta_\jmath^N$ in the conformal Newtonian gauge adopted for this calculation. 
$\delta_\jmath^N$ is defined on slices of constant coordinate time $\eta$, whence
\begin{equation}
a^3(z)\bar\jmath_\nu^\text{\tiny phys}(z)\Big(1+\delta_\jmath^z(z,\vx)\Big)
= a^3(\eta)\bar\jmath_\nu^\text{\tiny phys}(\eta)\Big(1+\delta_\jmath^N(\eta,\vx)\Big)
\end{equation}
Since, at fixed observed redshift, $\eta(z)=\bar\eta(z)+\delta\eta$ where $\delta\eta$ is given by 
Eq.\ (\ref{eq:deta}), we find after some algebra
\begin{equation}
\delta_\jmath^z = \delta_\jmath^N + 
\left(\frac{\dot{\bar\jmath}_\nu^\text{\tiny phys}}{\bar\jmath_\nu^\text{\tiny phys}}+3{\cal H}\right)
\delta\eta
= \delta_\jmath^N + \frac{\dot{\bar\jmath}_\nu}{\bar\jmath_\nu}\delta\eta
\end{equation}
at first order in perturbations.
Here and henceforth, $\jmath_\nu=a^3 \jmath_\nu^\text{\tiny phys}$ is the {\it comoving} emissivity.
Taking into account the factor $(1+z)^{-3}\equiv a^3(z)$ in Eq.\ (\ref{eq:Inuo_1}), the CIB brightness
simplifies to
\begin{equation}
\label{eq:Inuo_2}
I(\nu_o;\nvh) = \int\!\!\dd z\, \frac{a^2(z)}{{\cal H}(z)} \bar\jmath_\nu(z)
\left(1+ \delta_\jmath^N+\frac{\partial\ln\bar\jmath_\nu}{\partial\bar\eta}
\delta\eta + \delta_\parallel \right)\;,
\end{equation}
with $\nu=(1+z)\nu_o$. 

Lastly, we must also take into account the fact that the CIB is produced by sources below a certain flux 
limiting value $S_\nu^\text{\tiny cut}$. Owing to fluctuations in the luminosity distance, this corresponds 
to a maximum luminosity $L_\nu^\text{\tiny cut}(z,\nvh)$ which depends on redshift and position on the sky. 
At first order in perturbations, we get
\begin{equation}
L_\nu^\text{\tiny cut}(z,\nvh) = \bar L_\nu^\text{\tiny cut}(z) \big(1+2\delta_\perp\big) \;.
\end{equation}
The threshold luminosity $\bar L_\text{\tiny cut}(z)$ is related to the flux detection limit 
$S_\text{\tiny cut}$ through
\begin{equation}
\bar L_\nu^\text{\tiny cut}(z) = 4\pi \big(1+z\big)^4 a^2(z) \bar\chi^2(z) S_\nu^\text{\tiny cut} \;,
\end{equation}
where $\bar\chi(z)$ is the line-of-sight comoving distance corresponding to the redshift $z$ in the 
unperturbed background.
The perturbation $\delta_\perp$ transverse to the photon propagation at the source position is given by 
Eq.\ (\ref{eq:dperp}). Therefore, the average CIB emissivity $\bar\jmath_\nu$ 
per comoving volume in Eq.\ (\ref{eq:Inuo_2}) should be replaced by 
\begin{multline}
\int_0^{L_\nu^\text{\tiny cut}(z,\nvh)}\!\!\!\dd L \, \bar n_g(L,z) \frac{L_{(1+z)\nu}}{4\pi} \\
= \bar\jmath_\nu(z)
+ 2\, \bar n_g\big(\bar L_\nu^\text{\tiny cut}(z)\big) \frac{\bar L_\nu^\text{\tiny cut}(z)}{4\pi} 
\delta_\perp \;,
\nonumber
\end{multline}
where, in $\bar\jmath_\nu(z)$, the integral over galaxy luminosities runs from 0 to 
$\bar L_\nu^\text{\tiny cut}(z)$. The perturbed CIB comoving emissivity at hypersurfaces of constant observed
redshift thus reads
\begin{equation}
\label{eq:jmath2}
\jmath_\nu(z,\vx) = \bar\jmath_\nu(z)
\left(1+\delta_\jmath^N+\frac{\partial\ln\bar\jmath_\nu}{\partial\eta}\delta\eta
+\delta_\parallel
+2\frac{\bar n_g}{\bar\jmath_\nu}\frac{\bar L_\nu^\text{\tiny cut}}{4\pi}\delta_\perp\right)
\end{equation}
for a given flux detection limit $S_\nu^\text{\tiny cut}$.

Finally, all the background quantities are thus far been evaluated at $\bar\eta(z)$.
However, since the unperturbed photon path can also be parametrized as 
$x^\mu(\bar\chi)=(\eta_0-\bar\chi,\bar\chi)$, we can recast the CIB brightness into 
Eq.\ (\ref{eq:grCIB}) with the aid of $\dd\bar\chi/\dd z=a(\bar\eta(z))/{\cal H}(\bar\eta(z))$. 

\subsection{Imprint on the CIB angular power spectrum}
\label{app:cl}

Fluctuations in the CIB intensity can be expanded in the basis of spherical harmonics $Y_\ell^m$ as
\begin{equation}
I(\nu,\nvh)=\sum_{\ell m} a_{\ell m}(\nu) Y_\ell^m(\nvh) \;.
\end{equation}
The frequency-dependent multipoles $a_{\ell m}(\nu)$ are given by the line-of-sight integral
\begin{equation}
\label{eq:almnu}
a_{\ell m}(\nu) = \int\!\!\dd z\left(\frac{\dd\chi}{\dd z}\right)\, 
a(z) \bar\jmath_\nu(z) \, \Delta_{\ell m}(\nu,z) \;,
\end{equation}
where $\chi$ now is the unperturbed line-of-sight comoving distance (we drop hereafter the overline for
shorthand convenience), and
\begin{align}
\Delta_{\ell m}(\nu,z) &= \int\!\!\dd\nvh\, Y_\ell^{m\star}\!(\nvh)\,
\left(\delta_\jmath^N+ \frac{\dot{\bar\jmath}_\nu}{\bar\jmath_\nu}\delta\eta 
+ \delta_\parallel+2\frac{\bar n_g}{\bar\jmath_\nu}\frac{\bar L_\text{\tiny cut}}{4\pi}\delta_\perp\right) 
\nonumber \\
&=\sum_S \Delta_{\ell m}^S(\nu,z) 
\end{align}
has been schematically decomposed in \S\ref{sub:GRNG} into a sum of contributions induced by the 
first-order Newtonian gauge perturbations $\Psi$, $\Phi$, $v$ and the synchronous gauge density 
$\delta_m^\text{\tiny syn}$. Going to Fourier space and using the plane-wave expansion, some of the 
basic building blocks are, for instance, 
\begin{align}
\Delta_{\ell m}^\Phi 
&= \frac{i^\ell}{2\pi^2} \int\!\!\dd^3k\,T_\Phi(k,\chi)j_\ell(k\chi)Y_\ell^{m\star}\!(\kvh)\, \Phi_i(\vk) \\
\Delta_{\ell m}^{\vv\cdot\nvh} &=
\frac{i^\ell}{2\pi^2} \int\!\!\dd^3k\,T_v(k,\chi)j_\ell'(k\chi)Y_\ell^{m\star}\!(\kvh)\, \Phi_i(\vk) 
\nonumber \\
\Delta_{\ell m}^{\frac{\dd\vv}{\dd\eta}\cdot\nvh} &=
-\frac{i^\ell}{2\pi^2} \int\!\!\dd^3k\,k\,T_v(k,\chi)j_\ell''(k\chi)
Y_\ell^{m\star}\!(\kvh)\, \Phi_i(\vk) \nonumber \;.
\end{align}
In the second line, the prime denotes a derivative w.r.t. the argument of the spherical bessel function
$j_\ell(x)$. Total derivatives w.r.t. the unperturbed conformal time, $\dd/\dd\eta\equiv -\dd/\dd\chi$, 
bring an additional derivative w.r.t. the argument of $j_\ell(x)$.

Collecting all the terms and substituting into Eq.\ (\ref{eq:almnu}), we eventually arrive at
\begin{align}
a_{\ell m}(\nu) &= \frac{i^\ell}{2\pi^2}\int\!\!\dd^3k\!
\int\!\!\dd z\left(\frac{\dd\chi}{\dd z}\right)\,a(z)
\bar\jmath_\nu(z) \\
&\qquad \times 
F_\ell(\nu,k,z)\,Y_\ell^{m\star}\!(\kvh)\Phi_i(\vk) 
\nonumber \;.
\end{align}
The radial window function $F_\ell(\nu,k,z)$ sourced by the perturbations $\delta_\jmath^N$, $\delta\eta$
and $\delta_\parallel$ is given by Eq.\ (\ref{eq:Fl}).

For the lensing magnification effect, which is not included in the Fisher analysis as it is small, 
$F_\ell(\nu,k,z)$ reads
\begin{align}
F_\ell(\nu,k,&z) = \bigg\{j_\ell(k\chi)
\bigg[\left(\frac{1}{{\cal H}\chi}-1\right)T_\Psi - T_\Phi\bigg] 
+ \left(\frac{1}{{\cal H}\chi}-1\right) \nonumber \\
&\qquad\times
\int_0^\chi\!\!\dd\chi'\bigg[\big(T_\Psi+T_\Phi\big)+\frac{1}{\chi}
\big(\dot{T}_\Psi+\dot{T}_\Phi\big)\bigg]j_\ell(k\chi') \nonumber \\
&\quad + \frac{1}{2\chi}\int_0^\chi\!\!\dd\chi'\,\frac{\chi-\chi'}{\chi'}
\big(T_\Psi+T_\Phi\big) j_\ell(k\chi') \nonumber \\
&\quad + j_\ell'(k\chi) \left(1-\frac{1}{{\cal H}\chi}\right) T_v
\bigg\} 2\,s(\chi) \;,
\label{eq:Fl2}
\end{align}
with the magnification bias $s(\chi)$ given by Eq.\ (\ref{eq:magnibias}).

\bibliographystyle{mn2e}
\bibliography{references}

\label{lastpage}

\end{document}